\documentstyle[12pt,epsfig]{article}
\newlength{\dinwidth}
\newlength{\dinmargin}
\newcommand{\BE}{\begin{eqnarray}}
\newcommand{\EE}{\end{eqnarray}}

\newcommand{\ba}{\begin{array}}
\newcommand{\ea}{\end{array}}
\newcommand{\ben}{\begin{equation}}
\newcommand{\een}{\end{equation}}
\newcommand{\bea}{\begin{eqnarray}}
\newcommand{\eea}{\end{eqnarray}}
\newcommand{\gsim}{\mathrel{\mathop{\kern 0pt \rlap
   {\raise.2ex\hbox{$>$}}} \lower.9ex\hbox{\kern-.190em $\sim$}}}

\def\nn{\nonumber}

\def\e2{{\epsilon \over 2}}
\def\p{\prime}

\def\ttau{{\tilde \tau}}
\def\bz{{\bar z}}
\def\bw{{\bar w}}

\def\p{\partial}
\def\h{{1\over 2}}

\def\cM{{\cal M}}
\def\bT{{\bar T}}
\def\bx{{\bar x}}
\def\bu{{\bar u}}

\begin{document}
\thispagestyle{empty}
\addtocounter{page}{-1}
\begin{flushright}
UK-05-04\\
{\tt hep-th/0507080}
\end{flushright}
\vspace*{1.3cm} \centerline{\huge {\bf Branes wrapping Black Holes}}
\vspace*{1.2cm} \centerline{\bf Sumit R. Das${}^{a}$, Stefano
Giusto${}^{b}$, Samir D. Mathur${}^{b}$,} \vspace*{0.2cm}
\centerline{\bf Yogesh Srivastava${}^{b}$, Xinkai Wu${}^{a}$ and
Chengang Zhou${}^{a}$} \vspace*{0.8cm} \centerline{\it
${}^a$Department of Physics and Astronomy} \vspace*{0.2cm}
\centerline{\it University of Kentucky, Lexington, KY 40506, U.S.A.}
\vspace*{0.6cm} \centerline{\it ${}^b$Department of Physics}
\vspace*{0.2cm} \centerline{\it Ohio State University, Columbus, OH
43210, U.S.A. } \vspace*{1cm} \centerline{\tt
das@pa.uky.edu,giusto@pacific.mps.ohio-state.edu} \centerline{\tt
mathur@pacific.mps.ohio-state.edu, srivastava.21@osu.edu }
\centerline{\tt xinkaiwu@pa.uky.edu,czhou@pa.uky.edu}
\vspace*{1.5cm} \centerline{\bf Abstract} \vspace*{0.5cm}

We examine the dynamics of extended branes, carrying lower dimensional
brane charges, wrapping black holes and black hole microstates in M
and Type II string theory. We show that they have a universal
dispersion relation typical of threshold bound states with a total
energy equal to the sum of the contributions from the charges. In
near-horizon geometries of black holes, these are BPS states, and
the dispersion relation follows from supersymmetry as well as
properties of the conformal algebra. However they
break all supersymmetries of the full asymptotic geometries of black
holes and microstates. We comment on a recent proposal which
uses these states to explain black hole entropy.

\vspace*{1.1cm}

\baselineskip=18pt
\newpage

\def\ben{\begin{equation}}
\def\een{\end{equation}}
\def\be{\begin{equation}}
\def\ee{\end{equation}}
\def\bea{\begin{eqnarray}}
\def\eea{\end{eqnarray}}
\def\nn{\nonumber}

\section{Introduction}\label{intro}

The dynamics of stable extended branes in backgrounds containing
fluxes have played an important role in exploring non-perturbative
aspects of string theory. A particularly important class of such
objects are dielectric branes which are extended objects formed by a
collection of lower dimensional extended objects moving in a
transverse dimension via Myers' effect \cite{Myers:1999ps}. Dielectric
branes wrap {\em contractible} cycles in the space-time and therefore
do not carry any net charge appropriate to its dimensionality, but has
nonvanishing higher multipole moments. In a class of backgrounds
(e.g. $AdS$ space-times or their plane wave limits and certain D-brane
backgrounds) the energy due to the tension of the dielectric brane is
completely cancelled by the effect of the background flux, so that its
dispersion relation is that of a massless particle, which is why they
are called giant gravitons \cite{McGreevy:2000cw}-\cite{Das:2000ab}
. The energetics of such branes are usually determined by a classical
analysis : however these branes are BPS states which renders the
classical results exact.

Recently a different class of extended brane configurations have been
found in the near-horizon geometry of four dimensional extremal black
holes \cite{Simons:2004nm}-\cite{Gaiotto:2004ij} constructed e.g. from
intersecting $D4$ branes and some additional $D0$ charge.  The
near-horizon geometry is $AdS_2 \times S^2 \times K$ where $K$ is a
suitable six dimensional internal space (e.g. Calabi-Yau). These are
branes of various dimensionalities wrapping {\em non-contractible}
cycles of the compact directions.  The branes which are wrapped on
cycles in $K$ have a net charge in the full geometry and are similar
to giant gravitons - the tension of the brane is cancelled and one is
left with the dynamics of gravitons.  More interestingly, there are
BPS $D2$ branes wrapped on the $S^2$ with a worldvolume flux providing
a $D0$ brane charge, and posessing momentum along $K$. These branes do
not have net $D2$ charges in the full geometry - they only contribute
a net $D0$ charge.  The ground state is static in global time,
located at a radial coordinate determined by the $D0$ charge. These
configurations preserve half of the enhanced supersymmetries of the
near-horizon geometry, but {\em do not preserve any supersymmetry of
the full geometry}.

In \cite{Gaiotto:2004ij} it has been argued that such brane
configurations provide a natural understanding of the entropy of
the black hole background. The presence of a magnetic type flux
in the compact direction means that such a static brane carries
a nonzero momentum and is in fact in the lowest Landau level. This
means that the ground state is degenerate. It turns out that this
degeneracy is {\em independent of the $D0$ charge $q_0$ of the
background}. The idea then is to ``construct'' a black hole by
starting from the set of $D4$ branes and then add $D0$ charges.
However the $D0$ charge appear as these $D2$ branes which wrap
the $S^2$, and each such $D2$ has a ground state degeneracy.
The problem then reduces to a partitioning problem of distributing
a given $D0$ charge $N$ among $D2$ branes - the various possible
ways of doing this give rise to the entropy of the final black
hole. This argument has been extended to ``small'' black holes in
\cite{Kim:2005yb}.

In this paper, we show that such brane configurations are quite
generic not only in near-horizon geometries of black holes, but in the
{\em full} asymptotically flat geometries of certain black hole
microstates. While these are supersymmetric states in near-horizon
regions of black holes and near-cap regions of microstates, they break
all the supersymmetries of the asymptotically flat backgrounds.  We
find that in all cases they have a universal dispersion relation
characteristic of threshold bound states : the total energy is just
the sum of the energies due to various brane charges. In near-horizon
regions this simple dispersion relation follows from supersymmetry and
conformal algebras. However, we have not been able to find a good reason
why the same dispersion relation holds in the full microstate
geometries.

One key feature of the examples which we provide is that the
background does not have to posess the same kind of charge as the
brane itself. This feature could be relevant for the proposal of
\cite{Gaiotto:2004ij}, though we have reservations about this
proposal as it stands.

In section (\ref{general})
we consider generic $AdS_m \times S^n \times
{\cal M}$ space-times with a brane wrapped around $S^n$ and
moving along a $AdS$ direction with momentum $P$
and derive the universal
dispersion relation
\ben
E = P + M_n
\label{ione}
\een
where $M_n$ denotes the mass of the brane.

Specific examples of solutions of M theory and Type IIA string theory
which lead to $AdS_m \times S^n \times {\cal M}$ spacetimes are
described in section (\ref{examples}).  Our main example involves five
dimensional black strings in M theory compactified on $T^6$ (in
section (\ref{5D4D})) and their dimensional reduction to four
dimensional black holes in IIA theory (in section (\ref{IIABH})).  In
section (\ref{explain}) it is argued that the dispersion relation
(\ref{ione}) follows from the underlying conformal algebra. This is
explicitly shown for $AdS_3$, but the considerations should generalize
to other $AdS_m$. These branes are static in global time. In Poincare
time, they correspond to branes coming out of the horizon upto a
maximum distance and eventually returning back to the
horizon. However, we find that for $AdS_3$ (section (\ref{Poincare}))
and for $AdS_2$ (section (\ref{PoincareIIA}), the relation
(\ref{ione}) is valid {\em both in global and Poincare
coordinates}. Furthermore in $AdS_3$ the {\em Poincare momentum is
equal to the global momentum $P$}.  We argue that the equality of
global and Poincare energies and momenta signifies that the brane is
in a highest weight state of the conformal algebra.

The second class of backgrounds where we find such brane
configurations with identical dispersion relations are geometries
which represent microstate of 2-charge and 3-charge systems.  In the
examples of section (\ref{examples}) the existence of these brane
configurations appears to be special to near-horizon limits. This is
because they are states of lowest value of the global AdS energy and
not of the Poincare energy and it is the latter which coincides with
the energy defined in the full asymptotically flat geometry.  In
contrast, the microstate geometries are asymptotically flat and go
over to a {\em global patch} of $AdS$ in the interior.  The time in
the asymptotic region continues to the {\em global time} of the
interior $AdS$. Consequently, the notion of energy is unambigious.

In sections (\ref{microstate}) and (\ref{3chargemicro}) we find that
the lowest energy states of such branes are indeed static
configurations with dispersion relations given by (\ref{ione}).
Section (\ref{microstate}) deals with a T-dualized version of the
2-charge microstate geometry with D3 branes wrapping the $S^3$. We
show, in section (\ref{cftdual}) that the energy has an interesting
implication for the conformal field theory dual.  In section
(\ref{vibration}) we determine the spectrum of vibrations of the brane
and find a remarkably simple equispaced excitation spectrum with
spacing determined only by the AdS scale - reminiscent of the spectrum
of giant gravitons found in \cite{Das:2000st}. Section
(\ref{3chargemicro}) deals with analogous treatments of a special
3-charge microstate geometry.

In section (\ref{field}) we calculate the field produced by such
a probe brane in the 2-charge microstate geometry and show that this
leads to a {\em constant} field strength in the asymptotic region,
pretty much like a domain wall.

In section (\ref{susy}) we examine the supersymmetry properties of
these brane configurations. Section (\ref{susyd2}) deals with the case
of D2 branes in the background of 4d black holes, which is the
background of section (\ref{IIABH}). We show that in the near horizon
limit this D2 brane preserves half of the supersymmetries. We
calculate the topological charge on the brane and show that the
supersymmetry algebra leads to our simple dispersion relation. It is
then explicitly shown that the brane does not preserve any
supersymmetry of the full black hole geometry.  In section
(\ref{susymm}) we investigate the question in the 2 charge microstate
geometry and show that while the near-cap limit (which is again $AdS_3
\times S^3$) the brane preserves supersymmetry, it breaks all the
supersymmetries of the full background.

In an appendix we examine the validity of the
near-horizon approximation our brane trajectories for the
case of 4D black holes and show that the approximation is indeed
valid when the energy due to D0 charge of the D2 brane is smaller
than the D2 brane mass.

\section{Spherical branes $AdS \times S \times M$
space-times}\label{general}

The simplest space-times in which these brane configurations occur are
of the form $AdS_m \times S^n \times \cM$, where $\cM$ is some
internal manifold.

Let us first consider branes in M-theory backgrounds.
The metric is given by
\ben
ds^2 = R^2[-\cosh^2\chi~d\tau^2 + d\chi^2 + \sinh^2\chi~d\Omega_{m-2}^2]
+ {\tilde R}^2 d\Omega_n^2 + g_{ij}dy^i dy^j
\label{neone}
\een
where $R, {\tilde R}$ are length scales, $g_{ij}$ is the metric on $\cM$
and $d\Omega_p^2$ denotes the line
element on a unit $S^p$. We will choose coordinates $(\theta_k,\varphi)$
on $S^{m-2}$ leading to a metric
\ben
d\Omega^2_{m-2} = d\theta_1^2 + \sin^2\theta_1 ~d\theta_2^2
+ \sin^2 \theta_2 \sin^2\theta_1 ~d\theta_3^2 + \cdots
+ \sin^2 \theta_{m-3} \cdots \sin^2\theta_1~d\varphi^2
\label{netwo}
\een
The background could have  $m$-form and $n$ form gauge field
stengths which will not be relevant for our purposes.

In addition, the background contains
$(n+1)$-form gauge potentials ($n=2$ or $n=5$) of the form
\ben
A^{(n+1)} = A_i (y^i)~d\omega_n
\wedge d y^i
\een
where $d\omega_n$ denotes the volume form on the sphere. We will see
explicit examples of these geometries later.

Consider the motion of a $n$-brane which is wrapped on the $S^n$,
rotating in the $S^{m-2}$ contained in the $AdS_m$ and in general
moving along both $\chi$ and $y^i$.
The bosonic part of the brane action is of the form
\ben
S = -\mu_n \int d^{n+1}\xi~{\sqrt{{\rm det}~G}}
+\mu_n \int P[A^{(n+1)}]
\label{nethree}
\een
where $G$ denotes the induced metric, the symbol $P$ stands for pullback
to the worldvolume and $\mu_n$ is the tension of the $n$-brane.

Let us fix a static gauge where the worldvolume time is chosen to be
the target space time and the worldvolume angles are chosen to the
angles
on $S^n$. The remaining worldvolume fields are
$\chi,y^i,\theta_k,\varphi$.
When these fields are independent of the angles on the worldvolume,
the dynamics is that of a point particle. The Hamiltonian can
be easily seen to be
\ben
H = \cosh \chi {\sqrt{M_n^2 + {P_\chi^2 \over R^2}
+ {\Lambda^2 \over R^2 \sinh^2\chi} + g^{ij}(P_i-M_n A_i)(P_j- M_n A_j)}}
\label{nefour}
\een
where $M_n$ is the mass of the brane
\ben
M_n = \mu_n {\tilde R}^n \Omega_n
\een
$\Omega_n$ being the volume of unit $S^n$. $\Lambda$ denotes the conserved
angular momentum on $S^{m-2}$
\ben
\Lambda^2 = p_{\theta_1}^2 + {p_{\theta_2}^2 \over \sin^2\theta_1}
+ {p_{\theta_3}^2 \over \sin^2\theta_1 \sin^2\theta_2} + \cdots
{p_{\varphi}^2 \over \sin^2\theta_1 \cdots \sin^2\theta_{m-3}}
\een
Consider the lowest energy
state for some given $|\Lambda|$.
In the internal space this means that
$P_i = M_n A_i$. (This can be considered to be the description of  the lowest
Landau level in the classical limit). In $AdS$ this has
a fixed value of the global coordinate $\chi =
\chi_0$ determined by minimizing
the hamiltonian :
\ben
\sinh^2 \chi_0 =  {|\Lambda| \over R~M_n}
\label{nefive}
\een
The motion on the $S^{m-1}$ contained in $AdS_{m+1}$ is along an
orbit with
\ben
p_{\theta_k} = 0 ~~~\theta_k = {\pi \over 2}~~~~~k=1\cdots (m-3)
\een
The ground state energy is
\ben
E_{global} = {|\Lambda|\over R} + M_n
\label{nesix}
\een
Finally it is easy to check that in this state
\ben
{\dot{\varphi}} = 1
\een

While the above formulae have been given for M-branes, they
apply equally well for $D5$ branes in $AdS_5 \times S^5$
backgrounds of Type IIB string theory. This in fact
provides the simplest example of such configurations. We will give a
general explanation below for the simple form (\ref{nesix})
of the energy $E$.

\section{Extremal Black Strings in M theory and Black Holes
in String Theory}\label{examples}

In this section we will provide some concrete examples where
branes in $AdS \times S \times {\cal M}$ appear.

\subsection{5D Black Strings and 4D Black Holes}\label{5D4D}

A specific example of interest is the geometry of an extremal
black string in M-theory copmpactified on $T^6$ whose coordinates
are denoted by $y^1 \cdots y^6$.
The background is produced
by three sets of M5 branes which are wrapped on the directions
$y\,y^3\,y^4\,y^5\,y^6$,
$y\,y^1\,y^2\,y^5\,y^6$ and $y\,y^1\,y^2\,y^3\,y^4$ and carrying
momentum $q_0$ along $y$. The numbers $n_i$  and
charges $p_i$ of the M5 branes are related as
\bea
p_i={2\pi^2\, n_i\,\over  M_{11}^3\,T^{(i)}}\,,\,i=1,2,3
\label{neseven}
\eea
where $T^{(1)}$, $T^{(2)}$, $T^{(3)}$ are the volumes of
the 2-tori $(1,2)$, $(3,4)$ and $(5,6)$.

The metric and gauge fields produced by this system of branes is
\bea
ds^2&=&h^{-1/3}\,\Bigl[-dt^2+dy^2+{q_0\over r}\,(dt-dy)^2\Bigr]+
h^{2/3}\,[dr^2+r^2\,(d\theta^2+\sin^2\theta\,d\phi^2)]\nonumber\\
&+&h^{-1/3}\,\sum_{i=1,2,3}H_i\,ds^2_{T^{(i)}}
\label{neeight}
\eea
\ben
A^{(3)}=\sin\theta ~ d\theta d\phi\, [p_3~
{y^5 dy^6-y^6dy^5\over 2}+ p_2~
{y^3 dy^4-y^4 dy^3\over 2}+ p_1~ {y^1
dy^2-y^2 dy^1\over 2}]
\label{gfield}
\een
where we have defined
\bea
h=H_1\,H_2\,H_3\,,\quad H_i=1+{p_i\over r}\,,\,i=1,2,3\,,
\quad H_0=1+{q_0\over r}
\label{nenine}
\eea
$ds^2_{T^{(i)}}$ is the flat metric on the 2-torus of
volume $T^{(i)}$.

\subsubsection{Near-horizon limit with $q_0 = 0$}\label{q0ads3}

When $q_0 = 0$ the near-horizon limit is given by
$AdS_3\times S^2 \times T^6$. This may be seen
by re-defining coordinates
\ben
y = \lambda x ~~~~~~t = \lambda T~~~~~r= 4\lambda~u^2
\een
where we have defined
\ben
\lambda \equiv (p_1~p_2~p_3)^{1/3}
\een
Then for $r \ll p_i$ and $q_0=0$ the metric (\ref{neeight})
becomes
\bea
ds^2=(2{\lambda})^2\Bigl[{du^2\over u^2}+u^2\,
(-dT^2+dx^2)+{1\over 4}\,
(d\theta^2+\sin^2\theta\,d\phi^2)\Bigr]+
{1\over \lambda}\,\sum_{i=1,2,3}p_i\,ds^2_{T^{(i)}}
\label{neten}
\eea
which is
$AdS_3\times S^2 \times T^6$ in Poincare
coordinates.

One can further continue the metric to global
$AdS_3$ using the transformations
\bea
&& T
={\cosh\chi\,\sin\tau\over\cosh\chi\,\cos\tau-
\sinh\chi\,\sin\varphi}\,,\quad
x
={\sinh\chi\,\cos\varphi\over\cosh\chi\,\cos\tau-\sinh\chi\,
\sin\varphi}\nonumber\\
&& u =\cosh\chi\,\cos\tau-\sinh\chi\,\sin\varphi
\label{aeleven}
\eea
The resulting metric is
\bea
ds^2=(2{\lambda})^2\Bigl[d\chi^2-\sinh^2\chi\,d\tau^2+\cosh\chi^2\,
d\varphi^2+{1\over 4}\,(d\theta^2+\sin^2\theta\,d\phi^2)\Bigr]+{1\over
\lambda}\,\sum_{i=1,2,3}p_i\,ds^2_{T^{(i)}}
\label{11Dq0=0}
\eea

We can now consider a M2 brane wrapped around the $S^2$ and apply the
general results in equations (\ref{nethree})-(\ref{nesix}). For
some given momentum $P_\varphi$ in the $\varphi$ direction, the lowest
value of the global energy is given by
\ben
E_{gs} = P_\varphi + 8\pi \mu_2 \lambda^3
\een
which corresponds to a brane which is static in global time.

\subsubsection{Near-horizon limit with $q_0 \neq 0$}\label{qads3}

The near-horizon geometry for $q_0 \neq 0$ is again
$AdS_3\times S^2 \times T^6$. For $r \ll q_0,p_i$ we
have, from (\ref{neeight})
\bea
ds^2 & = & \lambda^2[\rho '(-dT'^2+dx'^2) + (dT'-dx')^2 +
{d u '^2 \over u '^2}] \nn \\
& + & \lambda^2 (d\theta^2+\sin^2\theta\,d\phi^2)
+ {1\over \lambda}\,\sum_{i=1,2,3}p_i\,ds^2_{T^{(i)}}
\label{neeleven}
\eea
where we have defined
\ben
y =({\lambda^3 \over q_0})^{1/2}~x'~~~~~~
t =({\lambda^3 \over q_0})^{1/2}~T'~~~~~~
r=q_0 u'
\een
With a further change of coordinates (\cite{ms})
\ben
\bT-\bx=e^{T'-x'}~~~~~~~
\bT+\bx=T'+x'+{2\over u'}~~~~~~
\bu = {{\sqrt{u'}}\over 2}~e^{-(T'-x')/2}
\een
the metric reduces to the standard form of the Poincare metric on
$AdS_3 \times S^2 \times T^6$
\bea
ds^2=(2 \lambda)^2\Bigl[{d\bu^2\over \bu^2}+\bu^2\,
(-d\bT^2+d\bx^2)+ {1\over 4}
(d\theta^2+\sin^2\theta\,d\phi^2)\Bigr]+
{1\over \lambda}\,\sum_{i=1,2,3}p_i\,ds^2_{T^{(i)}}
\label{netwelve}
\eea
which is identical to the metric (\ref{neten}). As before, one
can pass
to the global $AdS_3$ using the formulae above.

Thus we see that whether the background  has momentum in the $y$
direction or not
the near-horizon geometry  has the local form $AdS_3\times S^2\times
{\cal M}$, so the dynamics of the M2 brane will be similar
in the two cases.

\subsection{An explanation of the dispersion relation}\label{explain}

For branes moving in flat space, we expect that the total energy $E$
arises from the `rest energy'
$M$  and the momentum  $P$ by a relation of the type
$E=\sqrt{M^2+P^2}$. But for the
branes studied here we get a  linear relation of the type $E=P+M$.
   The momentum $P$ causes a
shift in radial position of the brane, where the redshift factor is
different, and in the end we end up
with this simple energy law.

As we will see in a later section the brane configuration considered
above is a BPS state which preserves half of the supersymmetries
of the background. The dispersion relation then follows from the
supersymmetry algebra.

It turns out that there is a simple derivation of this
linear relation for branes in
$AdS$ spacetime based on the bosonic part of the conformal algebra.
We will present
this for the case of $AdS_3 \times S^n$. We suspect that similar
considerations
would hold for arbitrary $AdS_{m}$.

A n-brane wrapped on $S^n$ becomes a point massive particle in
$AdS_3$. Its lagrangian
\be
L=-m[-{\p X^\mu\over \p\ttau} {\p X_\mu\over \p\ttau}]^{1\over 2}
\ee
where $m$ is the mass of the brane and ${\tilde{\tau}}$ denotes the
worldline parameter. The lagrangian
is invariant under the $SL(2,R)
\times SL(2,R)$ isometries of the background. Denoting the global
$AdS_3$ coordinates by $\tau,\chi,\varphi$ and defining
$z=\tau +\varphi, ~\bz=\tau-\varphi$ the generators are
\bea
L_0&=&i~\p_z\nonumber \\
L_{-1}&=&i~e^{-iz}~[{\cosh 2\chi\over \sinh 2\chi} \p_z -{1\over
\sinh 2\chi} \p_\bz+{i\over 2}
\p_\chi]\nonumber \\
L_{1}&=&i~e^{iz}~[{\cosh 2\chi\over \sinh 2\chi} \p_z -{1\over \sinh
2\chi} \p_\bz-{i\over 2}
\p_\chi]
\label{lone}
\eea
and
\bea
\bar L_0&=&i~\p_\bz\nonumber \\
\bar L_{-1}&=&i~e^{-i\bz}~[{\cosh 2\chi\over \sinh 2\chi} \p_\bz
-{1\over
\sinh 2\chi} \p_z+{i\over 2}
\p_\chi]\nonumber \\
\bar L_{1}&=&i~e^{i\bz}~[{\cosh 2\chi\over \sinh 2\chi} \p_\bz -{1\over
\sinh 2\chi} \p_z-{i\over 2}
\p_\chi]
\label{ltwo}
\eea
We have the algebra
\be
[L_0, L_{-1}]=L_{-1}, ~~[L_0, L_1]=-L_1, ~~[L_1, L_{-1}]=2 L_0
\ee
\be
[\bar L_0, \bar L_{-1}]=\bar L_{-1}, ~~[\bar L_0, \bar L_1]=-\bar L_1,
~~[\bar L_1, \bar L_{-1}]=2 \bar
L_0
\ee
The conserved quantities corresponding to these isometries are given
by the replacement
\be
-i\p_\mu~\rightarrow ~ P_\mu
\label{lthree}
\ee
in (\ref{lone}),(\ref{ltwo}).
The global coordinate energy $E_{global}$ and momentum $P_\varphi$
of the brane are related to the conserved
charges under
translations of
$t, \varphi$
\be
E_{global}=-P_\tau ~~~P=P_\varphi
\ee

Denote the parameter on the worldline of the particle by $\ttau$.
The kind of solution we have been considering is of the form
\be
\chi=\chi_0, ~~t=\ttau, ~~\varphi =\ttau
\ee
This is a geodesic in $AdS_3$.
The isometries of $AdS_3$ will move this to other geodesics.
The key property of our solution is that
\ben
{\bar z} = \tau - \varphi = {\rm constant}
\een
By a choice of the zero of $\tau$ we can choose this trajectory
to be along $\bz = 0$. On this trajectory the isometry
$\bar L_1-\bar
L_{-1}=-\p_\chi$ leads to a shift of the radial
coordinate $\chi$. Therefore applying this isometry transformation
we will get a new solution to the equations of motion of
the form
\be
\chi=\chi_0 + \epsilon ~~\tau =\ttau, ~~\varphi =\ttau
\label{lseven}
\ee
the meomenta conjugate to $z,\bz$ are
\be
P_z=\h(P_\tau + P_\varphi)=\h(P-E_{global}),
~~~P_\bz=\h(P_\tau-P_\varphi)=-\h(E_{global}+ P)
\ee
while the isometry $Q \equiv \bar L_1-\bar L_{-1}$ is given by
\be
Q=-e^{-i\bz}~[{\cosh 2\chi\over \sinh 2\chi} P_\bz -{1\over \sinh 2\chi}
P_z+{i\over 2}
P_\chi]
+~e^{i\bz}~[{\cosh 2\chi\over \sinh 2\chi} P_\bz -{1\over \sinh 2\chi}
P_z-{i\over 2}
P_\chi]
\ee
where we have used (\ref{lthree}).

We now observe that
\be
\{P_z, Q\}=0
\ee
so $P-E_{global}$ does not change under the shift. We thus see that for our
family of solutions given by (\ref{lseven}) we will have
\be
E_{global} = P+{\rm constant}
\ee
To fix the constant we can go to the geodesic at $\chi=0$ which has
$P=0$. Then we just get the energy
of the brane wrapped on the $S^n$, sitting at the center of
$AdS_3$, Calling this energy $M_n$,
we get
\be
E_{global} = P+M_n
\ee
giving  the simple additive relation between the mass and momentum
contributions to the energy.

\subsection{Poincare coordinate energies and momenta}\label{Poincare}

The brane discussed above is static in global coordinates and would
therefore correspond to a moving brane in Poincare time. In
this subsection we discuss some properties of dynamical quantities
in Poincare coordinates for branes in
$AdS_3 \times S^n$. The coordinate transformations are given in
equations (\ref{aeleven}).

A trajectory
$\chi = \chi_0,~\varphi = \tau$ becomes the
following trajectory in Poincare coordinates
\bea
x & = & \tanh \chi_0~(1+T \tanh \chi_0) \nn \\
u  & = & {\cosh \chi_0 \over
{\sqrt{T^2 (1+\tanh^2\chi_0) + 2T ~\tanh \chi_0 + 1}}}
\label{neeighteen}
\eea
Thus the brane pops out of the horizon $u=0$ at $T = -\infty$,
goes out to a maximum distance $u_{max}$ and returns back to
the horizon at $T = -\infty$. At the same time the coordinate
$x$ increases monotonically with $T$. The total elapsed proper
time is {\em finite}.
The value of $u_{max}$
can be calculated from the above trajectory and one gets
\ben
u_{max} = {\sqrt{\cosh~2\chi_0}}
\een
The Poincare energy is given by
\ben
E_{Poincare} = {M_n |g_{TT}| \over {\sqrt{|g_{TT}|-
g_{xx}{\dot x}^2 - g_{uu}{\dot u}^2}}}
= {M_n u^2 \over {\sqrt{u^2(1-{\dot x}^2)-{1\over u^2}{\dot u}^2}}}
\een
Using the value of $\chi_0$ in (\ref{nefive}) one finds that
\ben
E_{Poincare} = M_n~\cosh^2\chi_0 = {|L|\over R} + M_n = E_{global}
\een

In an analogous way one can verify that the momentum in global
coordinates, $P_\varphi$, equals the momentum in Poincar\`e
coordinates, $P_x$:
\bea
P_{x}=M_n\,{u^2\,{\dot x} \over
{\sqrt{u^2(1-{\dot x}^2)-{1\over u^2}{\dot u}^2}}}
=M_n\,\cosh^2\chi_0\,\tanh^2\chi_0= P_\varphi
\eea
where we
have used the fact that, for the above trajectory
$\dot{x}=\tanh^2\chi_0$.

The trajectory $\chi = \chi_0, \varphi = \tau$  clearly does not have
the
smallest possible value of $E_{Poincare}$.  The lowest value of
$E_{Poincare}$ is in fact zero and corresponds to the brane being
pushed to the horizon $u=0$.

The equality of global and Poincare energies can be
understood from the symmetries of AdS.
The generators of the $SL(2,R) \times SL(2,R)$ isometries of the
background have been given in global coordinates in equation
(\ref{lone}) and (\ref{ltwo}). The generators in Poincare
coordinates are given in terms of $w=T+x$ and $\bw = T-x$ by
\bea
H_{-1}&=&i\,\partial_{w}\nonumber\\
H_0&=&i\,\Bigl[w\,\partial_{w}-{u\over
2}\,\partial_{u}\Bigr]\nonumber\\
H_{1}&=&i\Bigl[w^2\,\partial_{w}-w\,u\,\partial_{u}-{1\over
u^2}\,\partial_{\bw}\Bigr]
\eea
and analogous ones with $H_i\to
{\bar H}_i$ and $w\to \bw$.

The relation between the two sets of generators is
\bea
H_0={L_1+L_{-1}\over 2}\,,\quad H_{\pm 1}=L_0\mp i\,{L_1-L_{-1}\over 2}
\eea
Since the global energy $E_{global}$ and the global momentum
$P_\varphi$ are equal to the Poincare energy $E_{Poincare}$ and the
Poincare momentum $P_x$ we must have
\bea
E_{global}=L_0+{\bar L}_0=E_{Poincare}=H_{-1}+{\bar H}_{-1}\,,\quad
P_\varphi=-L_0+{\bar L}_0=P_{x}=-H_{-1}+{\bar H}_{-1}
\eea
which implies
\ben
L_1 - L_{-1} = {\bar L}_1 - {\bar L}_{-1} = 0
\label{netwenty}
\een

The relations (\ref{netwenty}) may be readily verified for the
trajectory under question by calculating the corresponding
Noether charges. Computation of these charges require some care :
since the transformations involve time, we cannot compute
the charges starting from the static gauge lagrangian. Rather
we should compute this {\em before} we choose the worldvolume
time equal to the target space time. However we can choose
the worldvolume angles equal to the target space angles as before.
This partially gauge fixed lagrangian is given by
\bea
L=-{M_n\over 2}\,\sqrt{\dot{z}^2+\dot{\bz}^2+2 \cosh 2\chi\,
\dot{z}\,\dot{\bz}-(2\dot{\chi})^2}
\eea
where the dot denotes derivative with respect to the worldvolume time
$\tilde\tau$.
The Noether charges corresponding to the  $SL(2,R) \times SL(2,R)$
generators (\ref{lone}) and (\ref{ltwo}) are obtained by the
substitutions
\bea
-i\partial_z = P_z\,,\quad -i\partial_{\bar z} = P_{\bar z}\,,\quad
-i\partial_\chi=P_\chi
\eea
The momenta $P_z$, $P_{\bar z}$ and $P_\chi$ for a given configuration
are given by
\bea
P_z&=&-{M_n\over 2}\,{\dot{z}+\cosh2\chi\,\dot{\bar z}\over
\sqrt{\dot{z}^2+\dot{\bz}^2+2 \cosh 2\chi\,
\dot{z}\,\dot{\bz}-(2\dot{\chi})^2}}\nonumber\\
P_{\bar z}&=&-{M_n\over 2}\,{\dot{\bar z}+\cosh2\chi\,\dot{z}\over
\sqrt{\dot{z}^2+\dot{\bz}^2+2 \cosh 2\chi\,
\dot{z}\,\dot{\bz}-(2\dot{\chi})^2}}\nonumber\\
P_\chi&=&M_n\,{2\,\dot{\chi}\over \sqrt{\dot{z}^2+\dot{\bz}^2+2 \cosh
2\chi\,
\dot{z}\,\dot{\bz}-(2\dot{\chi})^2}}
\eea
For our configuration with $\chi=\chi_0$, $z=2\,{\tilde\tau}$, ${\bar
z}=0$ we find
\bea
P_z=-{M_n\over 2}\,\quad P_{\bar z}=-{M_n\over 2}\,\cosh2\chi_0\,,\quad
P_\chi=0
\eea
and thus the Noether charges evaluate to\footnote{Note that, for
$\chi_0\not=0$, our configuartion is not
symmetric under exchange of $z$ and $\bar z$: this is obviously because
we have chosen $\dot{\varphi}=1$. Another
solution can be obtained with the choice   $\dot{\varphi}=-1$.}
\bea
L_0&=&-P_z={M_n\over 2}\nonumber\\
L_{-1}&=&-e^{-iz}~[{\cosh 2\chi\over \sinh 2\chi} P_z -{1\over
\sinh 2\chi} P_\bz+{i\over 2}
P_\chi]=0\nonumber \\
L_{1}&=&-e^{iz}~[{\cosh 2\chi\over \sinh 2\chi} P_z -{1\over \sinh
2\chi} P_\bz-{i\over 2}
P_\chi]=0\nonumber\\
\bar L_0&=&-P_\bz={M_n\over 2}\,\cosh2\chi_0\nonumber \\
\bar L_{-1}&=&-e^{-i\bz}~[{\cosh 2\chi\over \sinh 2\chi} P_\bz -{1\over
\sinh 2\chi} P_z+{i\over 2}
P_\chi]={M_n\over 2}\,\sinh2\chi_0\nonumber \\
\bar L_{1}&=&-e^{i\bz}~[{\cosh 2\chi\over \sinh 2\chi} P_\bz -{1\over
\sinh 2\chi} P_z-{i\over 2}
P_\chi]={M_n\over 2}\,\sinh2\chi_0
\eea
 From the expressions above we verify that $L_1-L_{-1}=0$ and
${\bar L}_1-{\bar L}_{-1}=0$, which explains the equality of $E,P$
between the global and Poincare systems.
  We also note that
the charges satisfy the constraints
\bea
L_0^2-L_1\,L_{-1}= {\bar L}_0^2-{\bar L}_1\,{\bar L}_{-1}={M_n^2\over 4}
\eea
Further, note that $L_1=L_{-1}=0$, so the configuration is a highest
weight state of one of the $SL(2,R)$ aglebras. This gives
$L_0={M_n\over 2}$, which yields
$E=P+M_n$, the linear relation observed for the energy of the brane.

\subsection{Reduction to IIA Black Holes}\label{IIABH}

The geometry (\ref{neeight})-(\ref{nenine}) can be reduced
to IIA theory by a Kaluza Klein reduction along the $y$ direction.
Using the standard relation
\ben
ds_{11}^2 = e^{-2\Phi\over 3}ds_{10}^2 +e^{4\Phi\over 3}[dy - A_\mu
dx^\mu]^2
\label{neseventeen}
\een
where $ds_{10}^2$ is the string metric, $\Phi$ is the dilaton
and $A_\mu$ is the RR 1-form gauge field, it is straightforward
to see that we get
a 4-charge extremal black hole in four dimensions
\bea
ds^2 & = & -(H_0 h)^{-1/2} dt^2 + (H_0 h)^{1/2} [ dr^2 + r^2
d\Omega_2^2]
+ ({H_0 \over h})^{1/2}\sum_i H_i ds^2_{T_i}
\label{nethirteen} \nn \\
A^{(1)} & = & (1-{1\over H_0})~dt
\label{nefourteen} \nn \\
A^{(3)}& = & \sin\theta ~ d\theta d\phi\, [p_3~
{y^5 dy^6-y^6dy^5\over 2}+ p_2~
{y^3 dy^4-y^4 dy^3\over 2}+ p_1~ {y^1
dy^2-y^2 dy^1\over 2}] \nn \\
e^{\Phi} & = &  {H_0^3 \over h}
\label{nefifteen}
\eea

The near-horizon limit of this IIA metric depends on whether or
not $q_0$ is non-vanishing. For $q_0 = 0$ this has a null singularity
at $r=0$. Note that this limiting metric is {\em not} the
dimensional reduction of the metric (\ref{neten}).

For $q_0 \neq 0$ the geometry is $AdS_2 \times S^2
\times T^6$. This may be seen by looking at the above formulae
for $r \ll q_0, p_i$. The resulting metric, 1-form potential and
dilaton are given by
\bea
ds^2 & = & - {r^2\over R_{IIA}^2}dt^2 +  {R_{IIA}^2 \over r^2} dr^2 +
R_{IIA}^2(d\theta^2 + \sin^2\theta d\phi^2) \nn \\
& + &  {\sqrt{q_0p_1 \over p_2 p_3}}((dy^1)^2 + (dy^2)^2)
+ {\sqrt{q_0p_2 \over p_3 p_1}}((dy^3)^2 + (dy^4)^2)  \nn \\
& + & {\sqrt{q_0p_3 \over p_1 p_2}}((dy^5)^2 + (dy^6)^2) \nn \\
A^{(1)} & = & [1-{r \over q_0}]~dt \nn \\
A^{(3)}& = & \sin\theta ~ d\theta d\phi\, [p_3~
{y^5 dy^6-y^6dy^5\over 2}+ p_2~
{y^3 dy^4-y^4 dy^3\over 2}+ p_1~ {y^1
dy^2-y^2 dy^1\over 2}] \nn \\
e^{\Phi} & = & {q_0 \over R_{IIA}}
\label{aone2}
\eea
where
\ben
R_{IIA} = (q_0p_1p_2p_3p_4)^{1/4}
\label{atwo}
\een
If we replace the internal torus with a Calabi-Yau manifold,
this is the background which is used in \cite{Simons:2004nm}-
\cite{Gaiotto:2004ij}.

Equation (\ref{aone2}) is the metric in Poincare coordinates.
The coordinate transformations
\bea
{R_{IIA} \over r} & = & {1 \over \cosh \chi \cos \tau + \sinh \chi} \nn
\\
t & = & {R_{IIA} \cosh \chi \sin \tau \over \cosh \chi \cos \tau +
\sinh \chi}
\label{aten}
\eea
can be used to continue this metric to global coordinates
\bea
ds^2 & = & R_{IIA}^2(-\cosh^2\chi~d\tau^2 + d\chi^2) +
R_{IIA}^2(d\theta^2 + \sin^2\theta d\phi^2) \nn \\
& & + {\sqrt{q_0p_1 \over p_2 p_3}}((dy^1)^2 + (dy^2)^2)
+ {\sqrt{q_0p_2 \over p_3 p_1}}((dy^3)^2 + (dy^4)^2)  \nn \\
+ & & {\sqrt{q_0p_3 \over p_1 p_2}}((dy^5)^2 + (dy^6)^2)
\label{aone}
\eea
and one can choose a gauge in which the 1-form potential becomes
\ben
A^{(1)}  =  -{R_{IIA} \over q_0}[1- \sinh \chi] d\tau
\label{nesixteen}
\een

In the IIA language the M2 brane becomes a D2 brane and the
momentum along the $y$ direction becomes a D0 charge because
of the presence of a worldvolume gauge field
\ben
F = {f \over 2\pi \alpha^\prime}\sin \theta d\theta \wedge d\phi
\een
The contribution to the D0 brane charge to the mass of this
brane in string metric is
\ben
M_0 = 4\pi \mu_2 f
\een
where $\mu_2$ is the D2 brane tension. The global
hamiltonian may be
written down using standard methods
\ben
H =   \cosh \chi [(M_2^2 + M_0)^2~e^{-2\Phi} + P_\chi^2 +
{(P_i-4\pi\mu_2A_i)^2 \over g_{ii}}]^{1\over 2} +  M_0~e^{-\Phi}
[1- \sinh \chi]
\label{aeight}
\een
where in writing down the last term we have used the explicit
form of the dilaton in (\ref{aone2}). (Here $A^{(3)}\equiv
A_i~\cos \theta
~d\theta d\phi dy^i$.) We have also denoted the
mass of the D2 brane by $M_2$
\ben
M_2 = 4\pi R_{IIA}^2 \mu_2
\een
A static solution is obtained at a value of $\chi=\chi_0$ given by
\ben
\tanh \chi_0 = {M_0 \over {\sqrt{M_2^2 + M_0^2}}}
\een
and the value of the energy is
\ben
E = (M_0 + M_2)e^{-\Phi}
\een
which is what we expect from the dimensional reduction of the
M theory result.

Note that the magnitude of the energy depends on the gauge
choice for $A^{(1)}$. We have intentionally chosen a gauge
which leads to an energy which is identical to the M-theory result.
A gauge transformation on $A^{(1)}$ translates to a {\em coordinate}
transformation in the M theory which redefines the coorinate $y$
and therefore changes the Killing vector along which dimensional
reduction is performed to obtain the IIA theory. For example
instead of the choice in (\ref{aone2}) we could have chosen
\ben
A^{(1)\prime}  =  -{r \over q_0}~dt
\een
which is related to the original potential by a gauge transformation.
 From (\ref{neseventeen}) it is easy to see that this corresponds to
a coordinate transformation on $y$, $y \rightarrow y + t$. Thus
this gauge potential would arise from a KK reduction of the
11 dimensional metric along $y+t$ rather than $y$. In this situation
we do not of course expect the energy as calculated in IIA to agree
with the energy as calculated in M theory.

The expression for the hamiltonian, (\ref{aeight}) is not a sum
of positive terms and it is not evident that the static solution
has the lowest energy. However it is not hard to see that this is
indeed the ground state, using the trick of
\cite{Das:2000ab}. It is convenient to use coordinates
$\rho = \sinh \chi$ so that the metric of the $AdS$ part
becomes
\ben
ds^2 = -(1+\rho^2)~d\tau^2 + {d\rho^2 \over 1+\rho^2}
\een
The expression for the energy is
\ben
E = {{\sqrt{M_2^2 + M_0^2}} |g_{\tau\tau}| e^{-\Phi}
\over {\sqrt{|g_{\tau\tau}| - g_{\rho\rho}(\partial_\tau
\rho)^2}}} + M_0 e^{-\Phi} (1- \rho)
\een
This equation may be now re-written as
\ben
(\partial_\tau \rho)^2 + 2U(\rho) = 0
\een
where
\ben
2U(\rho) = {(M_0^2+M_2^2)(1+\rho^2)^3 \over ((Ee^{\Phi}-M_0) + M_0\rho)^2}
- (1+\rho^2)^2
\label{atwelve}
\een
The relativistic dynamics of the $D2$ brane is thus identical to the
{\em non-relativistic} dynamics of a particle of unit mass moving
in a potential $U(\rho)$. The energy of this analog non-reltivistic
problem
is zero.

A solution to this non-relativistic problem will exist only if
$U(\rho)=0$ for some real $\rho$. From (\ref{atwelve}) we see that
this happens
when
\ben
M_2^2\rho^2 - 2 M_0 (Ee^{\Phi}-M_0)~\rho -
(Ee^{\Phi}-M_0)^2+ (M_2^2+M_0^2) = 0
\een
This has a real solution only if
\ben
E \geq (M_2 + M_0)e^{-\Phi}
\een
which establishes the lower bound on the energy. When the energy
saturates this bound the solution is static.

\subsubsection{Poincare energies}\label{PoincareIIA}

The Poincare energies and momenta for this $D2$ brane are
again equal to the global energies and momenta. The
transformations are given in (\ref{aten}).
The trajectory is then given by
\ben \sinh \chi_0 = {u \over 2R_{IIA}}[1-({R_{IIA}^2 \over u^2} - {t^2
\over R_{IIA}^2})]
\label{aeleven2}
\een
This is again a trajectory which comes out of the horizon and
returns to it in finite proper time.
The maximum value of $u$ now turns out to be
\ben
u_{max} = R_{IIA}~e^{\chi_0}
\een
The value of the Poincare energy for this trajectory is
\ben
E_{Poincare} = {M g_{tt} e^{-\Phi}
\over {\sqrt{g_{tt}- g_{rr}(\partial_\tau
r)^2}}} +  M_0 e^{-\Phi}[1- {r \over R_{IIA}}]
= (M_0 + M_2)~e^{-\Phi}
\een
which is again {\em exactly} equal to the global energy $E_{global}$.

Just as in
the subsection (\ref{Poincare}), the
equality of Poincare and global energies has a group theoretic significance.
In terms of light cone coordinates
$t_\pm = t \pm {R^2_{IIA}\over r}$
the generators of the $SL(2,R)$ conformal isometries of $AdS_2$ are
\ben
h = L_{-1} = {\partial \over \partial t_+} + {\partial \over
\partial t_-},~~~~
d = L_{0}  =  t_+{\partial \over \partial t_+} +
t_-{\partial \over \partial t_-},~~
k = L_{1}  =  t_+^2{\partial \over \partial t_+} +
t_-^2{\partial \over \partial t_-}
\label{afourteen}
\een
and the transformation to global coordinates is given by
\ben
t_\pm = \tan ~[{1\over 2}(\tau \pm {1\over \cosh \chi})]
\een
The global hamiltonian $H$ is then
\ben
H_{global} = {\partial \over \partial \tau} = h + k
\een
Since the configurations we discussed have $H_{global} = h$ these must
have $k = 0$. $k$ is the generator of conformal boosts and the
standard $SL(2,R)$ algebra obeyed by $L_\pm,L_0$ then implies that
this state is a highest weight state.

The computations of these conserved charges follow the procedure
of subsection (\ref{Poincare}).
The partially gauge fixed action (for lowest Landau level orbits on the
$T^6$)
\ben
S = - {4\pi\mu_2 R_{IIA}^2 \over q_0}\int {d\tau \over v(\tau)}[
{\sqrt{R^4_{IIA}+f^2}}~
{\sqrt{(\partial_\tau t)^2 - (\partial_\tau v)^2}}
- f (\partial_\tau t)]
\een
The conserved charges in the static gauge are
\footnote{Note that the lagrangian is not
invariant under special conformal transformations, though the
action is - this results in an additional contribution to the
Noether charge}.
\bea
h & = & {4\pi\mu_2 R_{IIA}^2 \over q_0 v}[{A \over
{\sqrt{1 - {\dot{v}}^2}}} - f] \nn \\
d & = & {4\pi\mu_2 R_{IIA}^2 \over q_0 v}[-{A t\over
{\sqrt{1 - {\dot{v}}^2}}} + {A v {\dot{v}}\over
{\sqrt{1 - {\dot{v}}^2}}} + ft] \nn \\
k & = & {4\pi\mu_2 R_{IIA}^2 \over q_0 v}[ {A \over
{\sqrt{1 - {\dot{v}}^2}}}(tv{\dot{v}}-{1\over 2}(t^2+v^2))
-{f \over 2}(v^2-t^2)]
\label{afifteen}
\eea
Substituting the trajectory (\ref{aeleven2}) we find that $k$ evaluates
to zero.

\subsubsection{Validity of the near-horizon approximation}

The branes we discussed so far were shown to be stable and static in
global time in the near horizon geometry of the 4d extremal black
hole. From the point of view of black hole physics these would be of
interest only if they exist in the full asymptotically flat
geometry. In the full geometry, the near-horizon region is a
Poincare patch of $AdS$ and we have seen that in Poincare coordinates
the brane comes out of the horizon and goes back into it. This is what
one would expect in the full geometry as well. However we have to
check whether the approximation of restriction to the near-horizon
limit is self-consistent. In the Appendix, this is done for
four dimensional black hole geometry of section (\ref{IIABH}).
We find that the brane remains in the near-horizon region so long
as $M_0 \ll M_2$, but goes out of this region otherwise

\subsection{Examples in Type IIB String Theory}

Another example is provided by extremal black strings in Type IIB
string theory compactified on $T^4$ formed by two sets of
D3 branes intersecting along a line together with some momentum along the
intersection, and its dimensional reduction to five dimensional
black holes. The physics is identical to black strings in M theory
and their reduction to four dimensional black holes considered
above. The calculations are identical and will not be repeated here.

\section{$D3$ branes in 2-charge microstate
geometries}\label{microstate}

We have examined branes in $AdS_m\times S^n$ spaces, and computed their
energy. But we can make a symmetry
transformation in the AdS, and change what we  call $E$.  If on the
other hand we had an asymptotically flat spacetime then we
might get a physically unique definition of energy. Note also that the
goal of \cite{Gaiotto:2004ij} was to study black hole states. Black holes
have asymptotically flat geometries, and we measure the energy of
different excitations using the time at infinity. So it would be
helpful if we could study branes in spacetimes which have the {\em global}
$AdS_m\times S^n$ structure in some region (we will wrap the test
branes on the $S^n$) but which go over to asymptotically flat spacetime
at large $r$.

Interestingly, such geometries are given by microstates of the 2-charge
system. In \cite{bal, mm} it was found that metrics
carrying D1 and D5 charges and a certain amount of rotation had the
above mentioned property: they were asymptotically flat at
large $r$ but were $AdS_3\times S^3\times T^4$ in the small $r$ region.
The point to note is that the geometries were not just
{\it locally} $AdS_3\times S^3$ in the small $r$ region; rather the
small $r$ region had the shape of a `cap' which looked like the
region $r<r_0$ of {\it global} $AdS_3\times S^3$.

In detail, we take  type IIB string theory, compactified on
$T^4\times S^1$. We wrap D1 branes on the $S^1$  and we wrap D5 branes
on $S^1\times
T^4$.  Let the $S^1$ be parametrized by $y$, with $0<y<2\pi R$, and the
$T^4$ be parametrized by coordinates $y^1, y^2, y^3, y^4$
with a overall volume $V$.
For our present purposes we will do two T-dualities, in the directions
$y^1, y^2$, so that the system is composed of two sets of D3
branes. These branes extend along $y^1, y^2, y$ and along $y^3, y^4, y$
respectively. This does not change the nature of the
geometry that we have described above.

We can now consider a D3 brane wrapped over the $S^3$, and let it move
in the direction $y$. This situation with the D3 brane is
very similar to the case of the M2 brane that we had studied above, and
we expect to get similar results on the energy . But now
we can extend our analysis to a spacetime which is asymptotically flat,
so we can identify the charges which correspond to the
energy
$E$ (conjugate to time
$t$ at infinity) and the momentum $P$ (conjugate to the variable $y$).

We can extend the analysis to a class of geometries that carry {\it
three charges}: the two D3 brane charges as above as well as
momentum $P$ along $S^1$. The geometries for specific microstates of
this system were constructed in \cite{Giusto:2004id}, and these again
have an AdS type region at small $r$ and go over to flat space at
infinity.

In each of the above cases we find, somewhat surprisingly, that we
again get a relation of the form $E=P+Constant$.  This might
suggest that there is again an underlying symmetry that rotates orbits
of the wrapped brane, but we have not been able to
identify such a symmetry.

\subsection{The 2-charge microstate geometry}\label{MM}

The string frame metric is given by
\bea ds^2&=&-h^{-1}\,(dt^2-dy^2)+h f\,\Bigl({dr^2\over r^2+
a^2\,\gamma^2}+d\theta^2\Bigr) \nonumber\\
&+&h\Bigl(r^2 + {Q_1 Q_2 a^2\,\gamma^2\,
\cos^2\theta\over h^2 f^2 }\Bigr)\,\cos^2\theta\,d\psi^2 +h\Bigl(r^2 +
a^2\,\gamma^2 - {Q_1 Q_2 a^2\,\gamma^2 \sin^2\theta\over h^2 f^2
}\Bigr)\,\sin^2\theta\,d\phi^2 \nonumber\\
&-& 2\,{a\,\gamma\,\sqrt{Q_1
Q_2}\over h f}\,\cos^2\theta\,d\psi\,dy - 2\,{a\,\gamma\,\sqrt{Q_1
Q_2}\over h
f}\,\sin^2\theta\, d\phi\,dt\nonumber\\
&+&\sqrt{Q_2\over
Q_1}\,(dy_1^2 + dy_2^2)+\sqrt{Q_1\over Q_2}\,(dy_3^2 +
dy_4^2)\nonumber\\
h&=&\sqrt{\Bigl(1+{Q_1\over
f}\Bigr)\Bigl(1+{Q_2\over f}\Bigr)}\,,\quad f= r^2 + a^2\,\gamma^2
\cos^2\theta
\label{mm}
\eea
while the dilaton field vanishes. There is a 4-form potential given by
\bea
A^{(4)}&=&\Bigl[-{Q_1\over f+Q_1}\,dt\wedge dy-{Q_2\,(r^2+a^2\,\gamma^2
+Q_1)\over f+Q_1}\,\cos^2\theta\,d\psi\wedge
d\phi \nonumber\\
&-&{a\,\gamma\,\sqrt{Q_1 Q_2}\over f+Q_1}\,\cos^2\theta\,dt\wedge d\psi-
{a\,\gamma\,\sqrt{Q_1 Q_2}\over f+Q_1}\,\sin^2\theta\,dy\wedge d\phi
\Bigr]\wedge dy^1\wedge dy^2
\eea
However the experience of the previous sections show that the only role
of this is to put a probe $D3$ brane in a Lowest Landau level orbit on
the $T^4$. We will therefore ignore this in the following discussion.

This geomtery reduces to the asymptotically flat space-time
$M^{1,5}\times
T^4$ in the large $r$ limit. In the limit $r^2,a^2 \ll {\sqrt{Q_1Q_2}}$
the metric becomes
\bea
ds^2&=&\sqrt{Q_1 Q_2}\Bigl( {dr^2\over r^2+a^2\,\gamma^2}+{r^2\over Q_1
Q_2}\,dy^2 - {r^2+a^2\,\gamma^2\over Q_1 Q_2}\,dt^2\Bigr) \nonumber\\
&+&\sqrt{Q_1 Q_2}(d\theta^2 + \cos^2\theta\,d\psi'^2 +
\sin^2\theta\,d\phi'^2)+\sqrt{Q_2\over Q_1}\,(dy^2_1 +
dy^2_2)+\sqrt{Q_1\over Q_2}\,(dy^2_3 + dy^2_4)
\eea
where $\psi'$ and $\phi'$ are ``NS sector coordinates''
\be
\psi'=\psi-{a\,\gamma\over\sqrt{Q_1 Q_2}}\,y\,,\quad
\phi'=\phi-{a\,\gamma\over\sqrt{Q_1 Q_2}}\,t
\ee
For $\gamma=1$, this is precisely global $Ad_3 \times S^3 \times T^4$
as may be seen by making the coordinate transformations to
\ben
\tau  =  {a\gamma t \over {\sqrt{Q_1Q_2}}}~~~~~~~ \varphi  = {a~y\gamma \over {\sqrt{Q_1Q_2}}} ~~~~~~~
r  =  a \gamma~\sinh \chi
\label{relation}
\een
For $\gamma=1/k$, with $k$ integer
greater than 1, the
``near horizon'' geometry is an orbifod space of the type $(Ad_3 \times
S^3)/Z_k \times T^4$.

The geometry therefore smoothly interpolates between {\em global}
$AdS$ (or an orbifold of it) and flat space.  The key fact about this
geometry is that in the
small $r$ region $t$ is the {\em global} time in $AdS_3$, while in the
large
$r$ region the same $t$ is the usual Minkowski time in the
asymptoically flat space-time. This is in contrast to the geometry of
three charge black holes in five dimensions where the Minkowski time
of the asymptotic region becomes the {\em Poincare} time of the
near-horizon region. Therefore we can address the question of
wrapped $D3$ branes in the full geometry.

\subsection{$D3$ branes in 2-charge microstate geometry}

In the geometry described above, consider a D3 brane wrapping the
angular $S^3$ and carrying momentum $P$ along the circle $y$. This
brane couples to the background $F^{(5)}$ flux, which extends in the
$S^3$ directions as well as two of the directions of $T^4$, and hence
behaves like a charged particle moving in a magnetic field on
$T^4$. This system represents thus a five dimensional analogue of the
$S^2$ wrapped D2 brane in a 4d black hole, studied in section 2.

Choosing $t$, $\theta$, $\psi$ and $\phi$ as worldvolume coordinates,
the square root of the determinant of the metric induced on the D3
brane can be written in the form
\be
\sqrt{-{\rm det}
P(g)}=\sin\theta\,\cos\theta\,\sqrt{(r^2+a^2\,\gamma^2)\,F_1 -
r^2\,F_2\,\dot{y}^2}
\ee
where we have defined
\be F_1=r^2\,(f+Q_1+Q_2)+Q_1 Q_2 \,,\quad
F_2=(r^2+a^2\,\gamma^2)\,(f+Q_1+Q_2)+Q_1 Q_2
\ee

It is then straightforward to compute the  D3 brane Lagrangian
\be L= -\mu_3\,(2\pi)^2\,\int\!d\theta\,\sin\theta\,\cos\theta\,
\sqrt{(r^2+a^2\,\gamma^2)\,F_1 - r^2\,F_2\,\dot{y}^2}
\ee
the momentum conjugate to $y$
\be
P=\mu_3\,(2\pi)^2\,\int\!d\theta\,\sin\theta\,
\cos\theta\,{r^2\,F_2\,\dot{y}\over
\sqrt{(r^2+a^2\,\gamma^2)\,F_1 - r^2\,F_2\,\dot{y}^2}}
\label{p}
\ee
and the energy of the D3 brane
\be E=
\mu_3\,(2\pi)^2\,\int\!d\theta\,\sin\theta\,
\cos\theta\,{(r^2+a^2\,\gamma^2)\,F_1\over
\sqrt{(r^2+a^2\,\gamma^2)\,F_1 - r^2\,F_2\,\dot{y}^2}}
\label{e}
\ee

Though the $\theta$ integrals could be explicitly computed, we find
it more convenient to perform integrations only after having minimized
the energy.

The location at which the D3 brane stabilizes can be found by either
minimizing $E$ with respect to $r^2$ keeping $P$ fixed or minimizing
$L$ with respect to $r^2$ keeping $\dot{y}$ fixed. The second way is
the most convenient and yelds the following, surprinsingly simple,
result:
\bea
&&{\partial L\over \partial r^2}=0\,\Rightarrow\,
\partial_{r^2}[(r^2+a^2\,\gamma^2)F_1]- \partial_{r^2}[r^2
F_2]\,\dot{y}^2=0\\
&&\Rightarrow \dot{y}^2 = {r^2 (f+Q_1 +Q_2)+ Q_1
Q_2 + (r^2+a^2\,\gamma^2)(f+r^2 + Q_1 +Q_2)\over
(r^2+a^2\,\gamma^2)(f+Q_1 +Q_2) + Q_1 Q_2 + r^2 (f+r^2 +a^2\,\gamma^2
+ Q_1 +Q_2)}=1
\nonumber
\eea
The location at which the D3 brane sits
is then found by putting $\dot{y}=1$ in the expression (\ref{p}) for
$P$ and solving with respect to $r$. Note that for $\dot{y}=1$ the
square root which appears in the expression for $P$ and $E$ simplifies
\be \sqrt{(r^2+a^2\,\gamma^2)\,F_1 - r^2\,F_2} =
a\,\gamma\,\,\sqrt{Q_1 Q_2}
\ee
The expressions for the momentum and
energy of the D3 at its stable point are then
\bea P&=&
{\mu_3\,(2\pi)^2\over a\,\gamma\,\sqrt{Q_1
Q_2}}\,\int\!d\theta\,\sin\theta\,\cos\theta\,r^2 F_2\nonumber\\ &=&
{\mu_3\,(2\pi)^2\over 2\,a\,\gamma\,\sqrt{Q_1 Q_2}}\,r^2\,\Bigl[Q_1
Q_2 + (r^2+a^2\,\gamma^2)\, \Bigl( r^2 + {a^2\,\gamma^2\over 2} + Q_1
+ Q_2 \Bigr)\Bigr]\nonumber\\ E&=& {\mu_3\,(2\pi)^2\over
a\,\gamma\,\sqrt{Q_1
Q_2}}\,\int\!d\theta\,\sin\theta\,\cos\theta\,(r^2+a^2\,\gamma^2) F_1
\nonumber\\ &=& {\mu_3\,(2\pi)^2\over 2\,a\,\gamma\,\sqrt{Q_1
Q_2}}\,(r^2+a^2\,\gamma^2)\,\Bigl[Q_1 Q_2 + r^2 \, \Bigl( r^2 +
{a^2\,\gamma^2 \over 2} + Q_1 + Q_2 \Bigr)\Bigr]
\label{kone}
\eea
 From the expressions above we see that the dispersion relation of the
D3 brane is
\be
E=P+ 2\pi^2 \,\mu_3\,\sqrt{Q_1 Q_2}\,a\,\,\gamma
\label{ktwo}
\ee
Remarkably, this is {\em identical} to the formula we would
have obtained if we performed the analysis in the $AdS$ limit.
This may be easily seen
from the general formulae of section (\ref{general}) and
noting that the standard $AdS$ coordinates are related to
the coordinates $r,t,y$ by the equations in (\ref{relation})
and that the $AdS$ scale is given by $(Q_1Q_2)^{1/4}$.

We would like to emphasize that the definition of energy is
completely unambigious in this geometry because of the presence
of an asymptotically flat region. Furthermore from general grounds
we know that if we simply added pure momentum to the 2-charge
microstate geometry the additional ADM energy is simply equal to the
momentum. This is what happens if we take the formal limit $\mu_3 = 0$
in (\ref{ktwo}) which shows we have taken the zero of the energy
correctly.

\begin{figure}[ht]
\centerline{\epsfxsize=3.0in
\epsfysize=2.0in
    {\epsffile{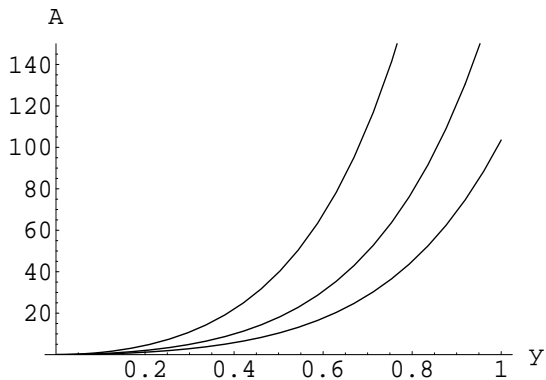}}
}
\caption{The ratio $A$ plotted as a function of $y$. The curves
have $b=0.1,0.15,0.2$ starting from top to bottom}
\label{mmgeom}
\end{figure}

Even though the dispersion relation is the same as in the $AdS$ limit,
the location of the brane obtained by solving the first equation of
(\ref{kone}) has a modified dependence on the momentum $P$. We would
like to determine the range of parameters for which this location lies
in the $AdS$ region. We have not obtained the general solution of the
equation. However to get an idea we examine the solution for $Q_1 = Q_2
= \lambda^2$. The quantity $\lambda$ will become the scale of the
$AdS$ in the appropriate region.
In this case it is useful to express this equation
in terms of the following quantities
\ben
A \equiv {P \over 2\pi^2 a \gamma \lambda^2 \mu_3}
~~~~y \equiv {r\over \lambda}
~~~~~~b={a \gamma \over \lambda}
\een
Note that $A$ is the ratio of the contributions to the from the momentum
and the $D3$ brane (as in (\ref{ktwo}). The $AdS$ region of the solution
corresponds to $y,b \ll 1$.

The first equation of (\ref{kone}) then becomes
\ben
A = ({y \over b})^2~[1+ (y^2+b^2)(y^2+{1\over 2}b^2 + 2)]
\een

Figure (\ref{mmgeom}) shows a plot of $A$ versus $y$ for various
values of $b$. The brane location moves further away from the center
of $AdS$ as we increase the ratio $A$, and for a given value of $A$,
the brane location $r=r_0$ is larger for larger values of $a$. This
shows that for small values of $b$ there is a large range of values of
$A$ for which the brane sits in the $AdS$ region of small $y$.

\subsection{CFT Duals}\label{cftdual}

In order to gain some insight on the dual CFT significance of the D3
brane configuarion discussed here, let us rewrite the expression above
in terms of microscopic quantities. If $R$ is the radius of the $y$
circle, $V=L_1\,L_2\,L_3\,L_4$ is the volume of $T^4$, $g$ the string
coupling and $n_1$ and $n_2$ are the numbers of D3 branes wrapped on
$y^1,y^2,y$ and $y^3,y^4,y$, one has
\be a={\sqrt{Q_1 Q_2}\over R}\,,\quad \mu_3={1\over
(2\pi)^3\,\alpha'^2\,g}\,,\quad Q_1={(2\pi)^2\,g\,\alpha'^2\over
L_3\,L_4}\,n_1\,,\quad Q_2={(2\pi)^2\,g\,\alpha'^2\over L_1\,L_2}\,n_2
\ee
and thus
\be E=P+2\pi^2\,\mu_3\,{Q_1 Q_2\over
k\,R}=P+{4\pi^3\,\alpha'^3\,g\over V}\,{1\over R}\,{n_1\,n_2\over k}
\ee

While the significance of this result is not clear to us, it s
interesting that the powers of the charges
are integral, so we get a quantity ${n_1n_2\over k}$ that counts the
number of `component strings' in the CFT microstate (see
\cite{lm} for a discussion of the microstate in terms of component
strings). Further the energy comes in units of ${1\over R}$
which is the natural qantum of energy in the CFT which lives on a
circle of radius $R$.

\subsection{Vibration modes}\label{vibration}

Let us look at the D3 brane cosidered above, and restrict attention to
the small $r$ region where the geometry is $AdS_3\times
S^3$. We have found the energy $E$ of the brane in a specific
configuration (which minimised $E$ for a given $P$), but we can
now ask for the properties of small vibratons of the brane around this
configuration.  We will only consider oscillations in the
$AdS_3$ directions, so that, in a static gauge, we can write
\be \chi=\chi_0+\epsilon\,\delta\chi(\tau,\theta_i)\,,\quad
y^5=\dot{y}\,\tau+\epsilon\,\delta y^5(\tau,\theta_i)\,,\quad
y^i=y_0^i\,,i=1\ldots,4
\ee
where we have denoted coordinates on $S^3$
by $\theta_i$, $i=1,\ldots,3$  and
the metric on a $S^3$ of unit radius by
$g_3$ .

We will compute the action of the D3 brane up to quadratic order in
$\epsilon$. Having suppressed oscillations in
the $T^4$ directions, only the DBI term contributes. The term of first
order in $\epsilon$ is
\bea
S_1=\epsilon\,\mu_3\,\lambda^4\,\int d\tau\,d^3\theta_i\,\sqrt{g_3}\,
{\sinh\chi_0\over \sqrt{\cosh^2\chi_0
-\dot{y}^2\,\sinh^2\chi_0}}\,[(\dot{y}^2-1)
\,\cosh\chi\,\delta\chi+\dot{y}\,\sinh\chi_0\,\partial_\tau\delta y^5]
\eea
The term proportional to $\delta\chi$ vanishes for $\dot{y}^2=1$, while
the coefficient of $\partial_\tau\delta y^5$
is a constant and thus this term does not contribute to the equations
of motion. Restricting to  $\dot{y}^2=1$, and
performing the change of coordinates $\rho=\sinh\chi$, the term of
second order in $\epsilon$ is
\bea
\!\!\!\!\!\!\!\!\!\!\!\!\!\
S_2&=&-\epsilon^2\,\mu_3\,\lambda^4\,\int
d\tau\,d^3\theta_i\,\sqrt{g_3}\,\Bigl[{g^{ij}_3\over 2}\,
{\partial_i\delta\rho\,\partial_j\delta\rho\over \rho_0^2+1}-{1\over
2}\,{(\partial_t\delta\rho)^2\over \rho_0^2+1}
\nonumber\\
&&\qquad +{g^{ij}_3\over 2}\,\rho_0^2(\rho_0^2+1)\,\partial_i\delta y^5
\,\partial_j\delta y^5 -{1\over
2}\,\rho_0^2(\rho_0^2+1)\,(\partial_t\delta y^5)^2
-2\,\rho_0\,\delta\rho\,\partial_t\delta y^5\Bigr]
\label{quadratic}
\eea

If one expands the perturbations $\delta\rho$ and $\delta y^5$ as
\be
\delta\rho(\tau,\theta_i)=\delta{\tilde\rho}\,e^{-
i\omega\tau}\,Y_l(\theta_i)\,,\quad
\delta y^5(\tau,\theta_i)=\delta{\tilde
y^5}\,e^{-i\omega\tau}\,Y_l(\theta_i)
\ee
where $Y_l$ are spherical harmonics on $S^3$
\be
{1\over \sqrt{g_3}}\,\partial_i\,( g_3^{ij} \partial_j
Y_l(\theta_i))=-Q_l\,Y_l(\theta_i)\,,\quad Q_l=l(l+2)
\ee
the equations of motion derived from the action (\ref{quadratic}) become
\be
\left(\begin{array}{clc}(\rho_0^2+1)^{-1}\,(-Q_l+\omega^2)& -2
\,i\,\omega\,\rho_0\\
2\, i\,\omega\,\rho_0 &
\rho_0^2\,(\rho_0^2+1)\,(-Q_l+\omega^2)\end{array}\right)=
\left(\begin{array}{c}\delta\tilde\rho\\
\delta\tilde y^5\end{array}\right)
\ee
The vibration frequencies are then
\be
\omega^2=Q_l+2\pm 2\sqrt{Q_l+1}=l(l+2)+2\pm 2 (l+1)
\ee
or equivalently
\be
\omega=l+2\quad {\rm and}\quad \omega = l
\ee
Note that $\omega$ denotes the conjugate of the dimensionless
coordinate $\tau$. This is related to the physical energies
by a suitable factor of the $AdS$ scale.
We therefore see that {\em the frequencies are
universal}. They
depend only on the $AdS$ scale of the background and not on the
value of the momentum $P$ of the brane. This is similar to
what happens for giant gravitons \cite{Das:2000st}.

\section{D3 branes in 3-charge microstates}\label{3chargemicro}

By applying a spectral flow to the two charge microstate of the
previous subsections one obtains a geometry dual to a three charge
microstate. This is described, in the string frame, by the following
metric and dilaton \cite{Giusto:2004id}
  \bea
\label{3charge}
ds^2 & = & -\frac{1}{h} (dt^2-dy^2) + \frac{Q_{p}}{h
f}\left(dt-dy\right)^{2}+ h f \left( \frac{dr^2}{r^2 +
(\gamma_1+\gamma_2)^2\eta} + d\theta^2 \right)\\
       &+& h \Bigl( r^2 + \gamma_1\,(\gamma_1+\gamma_2)\,\eta -
\frac{Q_1 Q_2\,(\gamma_1^2-\gamma_2^2)\,\eta\,\cos^2\theta}{h^2 f^2}
\Bigr)
\cos^2\theta d\psi^2  \nonumber \\
&+& h\Bigl( r^2 + \gamma_2\,(\gamma_1+\gamma_2)\,\eta +
\frac{Q_1 Q_2\,(\gamma_1^2-\gamma_2^2)\,\eta\,\sin^2\theta}{h^{2} f^{2}
}
\Bigr) \sin^2\theta d\phi^2  \nonumber \\
&+& \frac{Q_p\,(\gamma_1+\gamma_2)^2\,\eta^2}{h f}
\left( \cos^2\theta d\psi + \sin^2\theta d\phi \right)^{2} \nonumber\\
&-& \frac{2 \sqrt{Q_{1}Q_{2}} }{hf}
\left(\gamma_1 \cos^2\theta d\psi + \gamma_2 \sin^2\theta d\phi\right)
(dt-dy)
\nonumber \\
&-& \frac{2 \sqrt{Q_1 Q_2}\,(\gamma_1+\gamma_2)\,\eta}{h f}
\left( \cos^2\theta d\psi + \sin^2\theta d\phi \right) dy +
\sqrt{H_1\over H_2}(dy_1^2+dy_2^2)+\sqrt{H_2\over
H_1}(dy_3^2+dy_4^2)\nonumber
\eea

\be
e^{2\Phi}= \frac{H_{1}}{H_{2}}
\ee
where
\bea
&&\!\!\!\!\!\!\!\!\!\!\!\!\eta = {Q_1 Q_2\over Q_1 Q_2 + Q_1 Q_p + Q_2
Q_p}
\nonumber\\
&&\!\!\!\!\!\!\!\!\!\!\!\!f = r^2+ (\gamma_1+\gamma_2)\,\eta\,
\bigl(\gamma_1\, \sin^2\theta + \gamma_2\,\cos^2\theta\bigr) \nonumber\\
&&\!\!\!\!\!\!\!\!\!\!\!\!H_{1} = 1+
\frac{Q_{1}}{f}\,,\quad H_{2} = 1+ \frac{Q_{2}}{f}\,,\quad h =
\sqrt{H_{1} H_{2}}
\label{deffh}
\eea

For the solution obtained by spectral flow from the 2-charge microstate
geometry, the
parameters $\gamma_1$ and $\gamma_2$ take the values
\be
\gamma_1=-a\,n\,,\quad \gamma_2 = a\,\Bigl(n+{1\over k}\Bigr)\,,\quad
a={\sqrt{Q_1 Q_2}\over R}
\label{sf}
\ee
where $R$ is the $y$ radius and $n$ and $k$ are integers. Geometries
corresponding to other values
of $\gamma_1$ and $\gamma_2$ can be obtained by $S$ and $T$ dualities.

In this geometry, consider a D3 brane wrapping the angular $S^3$ and
rotating along $y$.
The determinant of the induced metric in static gauge can be cast the
the form
\be
\sqrt{-{\rm
det}P(g)}=-\sin\theta\cos\theta\,\sqrt{c_0+\dot{y}\,c_1+\dot{y}^2\,c_2}
\ee
where $c_0$, $c_1$ and $c_2$ are functions of $r$ and $\theta$ that can
be computed using Mathematica.
As we did not manage to bring these functions to reasonably simple
form, we do not give their explicit
expressions here. We can however proceed with the help of Mathematica
and  verify that the $r$-derivative of the
Lagrangian
\be
L=
-
\mu_3\,(2\pi)^2\,\int\!d\theta\,\sin\theta\,\cos\theta\,\sqrt{c_0+\dot{y
}\,c_1+\dot{y}^2\,c_2}
\ee
at fixed $\dot{y}$ vanishes for $\dot{y}=1$ (note that in this case the
invariance under $y\to-y$ is broken
by the momentum carried by the background metric (\ref{3charge}) and
$\dot{y}=-1$ is not a local
minimum). For this value of $\dot{y}$ the determinant of the induced
metric simplifies to
\be
\sqrt{-{\rm
det}P(g)}=-\sin\theta\cos\theta\,(\gamma_1+\gamma_2)\,\eta\,\sqrt{Q_1
Q_2}
\ee

Following the same steps as in the previous subsection, one can compute
the energy and momentum conjugate to $y$
at the stable point $\dot{y}=1$:
\bea
E&=&{\mu_3\,(2\pi)^2\over (\gamma_1+\gamma_2)\,\eta\,\sqrt{Q_1
Q_2}}\,\int d\theta\,\sin\theta\,\cos\theta\,{2c_0+c_1\over
2}\nonumber\\
P&=&-{\mu_3\,(2\pi)^2\over (\gamma_1+\gamma_2)\,\eta\,\sqrt{Q_1
Q_2}}\,\int d\theta\,\sin\theta\,\cos\theta\,{2c_2+c_1\over 2}
\eea
Neither $E$ or $P$ have particuarly good looking expressions, but their
difference is simply given by
\be
E=P+2\pi^2\,\mu_3\,\sqrt{Q_1
Q_2}\,(\gamma_1+\gamma_2)\,\eta= P + 2\pi^2\,\mu_3\,{\sqrt{Q_1 Q_2}\,a\over
  k}\,
\eta
\ee
where in the last equality we have used the values (\ref{sf}) for
$\gamma_1$ and $\gamma_2$.

We thus conclude that the dispersion relation of the D3 brane in the
three charge geometry differs from that in the
two charge geometry only by a factor of $\eta$.

\section{The field produced by the wrapped brane}\label{field}

In this section we look at the gauge field produced by the D3 brane
that we wrap on the $S^3$, in the asymptotically flat 2-charge
microstate geometry.  If we think of the brane as a small perturbation
of strength $\epsilon$ on the background, then the field
strength produced by the brane is also of order $\epsilon$, and the
energy carried by this field is $O(\epsilon^2)$. But we find
that the field strength goes to a constant at large
$r$, so that its overall energy would diverge. The brane wrapped on the
$S^3$ appears to behave like a domain wall in the
spacetime, making the field nonzero on the outside everywhere.

The action for the 4-form RR field $A^{(4)}$ sourced by the D3 brane is
\ben
S={1\over 2}\,\int F^{(5)}\wedge \star F^{(5)}+\mu_2\,\int dr\, dy\,
dt\,
d\theta\, d\phi\, d\psi\,dV\,\delta(r-r_0)\,
[A^{(4)}_{t\theta\phi\psi} +A^{(4)}_{y\theta\phi\psi}]
\een
($dV=dy^1\wedge\ldots\wedge dy^4$ is the volume form on $T^4$). We have
assumed the brane to be
smeared along $y$ and the torus directions $y^i$ and,
in writing the source term, we have taken into account that the brane
moves with velocity
$\dot{y}=1$ along $y$.

We will make the following ansatz for $A^{(4)}$
\bea
A^{(4)}&=&A^{(4)}_{t\theta\phi\psi}\,dt\wedge d\theta\wedge d\phi\wedge
d\psi+ A^{(4)}_{t\theta\phi y}\,
dt\wedge d\theta\wedge d\phi\wedge dy\nonumber\\
&+&A^{(4)}_{y\theta\phi\psi}\,dy\wedge d\theta\wedge d\phi\wedge d\psi+
A^{(4)}_{y\theta t \psi}\,
dy\wedge d\theta\wedge d t\wedge d\psi
\label{ansatz}
\eea
(At the same order in $\mu_2$, the gauge field also has components
$A^{(4)}_{\mu\nu y^1 y^2}$, where
$\mu,\nu=t,y,\psi,\phi$ and $y^1$, $y^2$ are directions in $T^4$: these
components arise from the fact that
the background metric is perturbed by the D3 brane together with the
fact
that the unperturbed background has
non-zero values of $A^{(4)}_{\mu\nu y^1 y^2}$. Since the equations of
motion do not mix the components
$A^{(4)}_{\mu\nu y^1 y^2}$ with the ones contained in the ansatz
(\ref{ansatz}), we can consistently ignore
these extra components in the following).

Ona has
\ben
F^{(5)}=dr\wedge \partial_r A^{(4)}
\een

The star operation in a geometry with $t\phi$ and $y\psi$ mixings is
given
by\footnote{We are using the orientation
$\epsilon_{tyr\theta\phi\psi}=1$.}
\bea
&&\star\,(dr\wedge dt\wedge d\theta\wedge d\phi\wedge d\psi)=
\sqrt{-g}\,g^{rr}\,g^{\theta\theta}\,(g^{tt}g^{\phi\phi}-
g^{t\phi}g^{t\phi})\,(g^{\psi\psi}\,dy-g^{\psi
y}\,d\psi)
\wedge dV\nonumber\\
&&\star\,(dr\wedge dt\wedge d\theta\wedge d\phi\wedge d y)=
\sqrt{-g}\,g^{rr}\,g^{\theta\theta}\,(g^{tt}g^{\phi\phi}-
g^{t\phi}g^{t\phi})\,(g^{\psi
y}\,dy-g^{y y}\,d\psi)
\wedge dV\nonumber\\
&&\star\,(dr\wedge dy\wedge d\theta\wedge d\phi\wedge d\psi)=
-\sqrt{-g}\,g^{rr}\,g^{\theta\theta}\,(g^{yy}g^{\psi\psi}-
g^{y\psi}g^{y\psi})\,(g^{\phi\phi}\,dt-g^{\phi
t}\,d\phi)
\wedge dV\nonumber\\
&&\star\,(dr\wedge dy \wedge d\theta\wedge d t\wedge d\psi)=
-\sqrt{-g}\,g^{rr}\,g^{\theta\theta}\,(g^{yy}g^{\psi\psi}-
g^{y\psi}g^{y\psi})\,(g^{\phi
t}\,dt-g^{t t}\,d\phi)
\wedge dV
\eea

The equations of motion are
\bea
d\star F^{(5)}+\mu_2\,\delta(r-r_0)\,dr\wedge (dy-dt)\wedge dV=0
\eea
which yeld
\bea
&&\partial_r[\sqrt{-g}\,g^{rr}\,g^{\theta\theta}\,
(g^{tt}g^{\phi\phi}-g^{t\phi}g^{t\phi})\,(g^{\psi\psi}\partial_r
A^{(4)}_{t\theta\phi\psi}+
g^{\psi y}\partial_r A^{(4)}_{t\theta\phi y})
]+\mu_2\,\delta(r-r_0)=0\nonumber\\
&&\partial_r[\sqrt{-g}\,g^{rr}\,g^{\theta\theta}\,
(g^{tt}g^{\phi\phi}-g^{t\phi}g^{t\phi})\,(g^{y y}\partial_r
A^{(4)}_{t\theta\phi y}+
g^{\psi y}\partial_r A^{(4)}_{t\theta\phi \psi})
]=0\nonumber\\
&&\partial_r[\sqrt{-g}\,g^{rr}\,g^{\theta\theta}\,
(g^{yy}g^{\psi\psi}-g^{y\psi}g^{y\psi})\,(g^{\phi\phi}\partial_r
A^{(4)}_{y\theta\phi\psi}+
g^{\phi t}\partial_r A^{(4)}_{y\theta t \psi})
]+\mu_2\,\delta(r-r_0)=0\nonumber\\
&&\partial_r[\sqrt{-g}\,g^{rr}\,g^{\theta\theta}\,
(g^{yy}g^{\psi\psi}-g^{y\psi}g^{y\psi})\,(g^{t t}\partial_r
A^{(4)}_{y\theta t\psi}+
g^{\phi t}\partial_r A^{(4)}_{y\theta \phi \psi})
]=0
\eea
\bea
&&\partial_\theta[\sqrt{-g}\,g^{rr}\,g^{\theta\theta}\,
(g^{tt}g^{\phi\phi}-g^{t\phi}g^{t\phi})\,(g^{\psi\psi}\partial_r
A^{(4)}_{t\theta\phi\psi}+
g^{\psi y}\partial_r A^{(4)}_{t\theta\phi y})
]=0\nonumber\\
&&\partial_\theta[\sqrt{-g}\,g^{rr}\,g^{\theta\theta}\,
(g^{tt}g^{\phi\phi}-g^{t\phi}g^{t\phi})\,(g^{y y}\partial_r
A^{(4)}_{t\theta\phi y}+
g^{\psi y}\partial_r A^{(4)}_{t\theta\phi \psi})
]=0\nonumber\\
&&\partial_\theta[\sqrt{-g}\,g^{rr}\,g^{\theta\theta}\,
(g^{yy}g^{\psi\psi}-g^{y\psi}g^{y\psi})\,(g^{\phi\phi}\partial_r
A^{(4)}_{y\theta\phi\psi}+
g^{\phi t}\partial_r A^{(4)}_{y\theta t \psi})
]=0\nonumber\\
&&\partial_\theta[\sqrt{-g}\,g^{rr}\,g^{\theta\theta}\,
(g^{yy}g^{\psi\psi}-g^{y\psi}g^{y\psi})\,(g^{t t}\partial_r
A^{(4)}_{y\theta t\psi}+
g^{\phi t}\partial_r A^{(4)}_{y\theta \phi \psi})
]=0
\eea
Their solution is
\bea
&&F^{(5)}_{rt\theta\phi\psi}={a_\pm\,g_{\psi\psi}+b_\pm\,g_{\psi y}\over
\sqrt{-g}\,g^{rr}\,g^{\theta\theta}\,
(g^{tt}g^{\phi\phi}-g^{t\phi}g^{t\phi})}\nonumber\\
&&F^{(5)}_{rt\theta\phi y}={a_\pm\,g_{\psi y}+b_\pm\,g_{yy}\over
\sqrt{-g}\,g^{rr}\,g^{\theta\theta}\,
(g^{tt}g^{\phi\phi}-g^{t\phi}g^{t\phi})}\nonumber\\
&&F^{(5)}_{ry\theta\phi\psi}={c_{\pm}\,g_{\phi\phi}+d_\pm\,g_{\phi
t}\over
\sqrt{-g}\,g^{rr}\,g^{\theta\theta}\,
(g^{yy}g^{\psi\psi}-g^{y\psi}g^{y\psi})}\nonumber\\
&&F^{(5)}_{ry\theta t\psi}={c_{\pm}\,g_{\phi t}+d_\pm\,g_{t t}\over
\sqrt{-g}\,g^{rr}\,g^{\theta\theta}\,
(g^{yy}g^{\psi\psi}-g^{y\psi}g^{y\psi})}
\eea
where $a_\pm$, $b_\pm$, $c_\pm$ and $d_\pm$ are $r$ and $\theta$
independent constants: the subscript $+$ applies
  to the region $r>r_0$ while  the subscript $-$
applies to $r<r_0$. Because of the delta function source we have
$a_+-a_-=-\mu_2$, $b_+-b_-=0$,
$c_+-c_-=-\mu_2$, $d_+-d_-=0$.

In order to fix the values of these constants let us impose regularity
of
the field strength.
It will be convenient to work in ``NS-sector coordinates''
\bea
\phi'=\phi-{a\over \sqrt{Q_1 Q_2}}\,t\,,\quad \psi'=\psi-{a\over
\sqrt{Q_1 Q_2}}\,y
\eea

Consider first regularity at $\theta=0,\pi/2$. One has
\bea
&&\sqrt{-g}\,g^{rr}\,g^{\theta\theta}\,(g^{tt}g^{\phi'\phi'}-
g^{t\phi'}g^{t\phi'})=-{r\over
h f}\,
{\cos\theta\over \sin\theta}\nonumber\\
&&\sqrt{-g}\,g^{rr}\,g^{\theta\theta}\,(g^{yy}g^{\psi'\psi'}-
g^{y\psi'}g^{y\psi'})={r^2+a^2\over
r\,h f}\,{\sin\theta\over \cos\theta}
\eea
Moreover $g_{\phi'\phi'}\sim\sin^2\theta$, $g_{t\phi'}\sim\sin^2\theta$,
$g_{\psi'\psi'}\sim\cos^2\theta$,
$g_{y\psi'}\sim\cos^2\theta$ while $g_{tt}$ and $g_{yy}$ go to some
finite
non-zero values as $\theta\to 0,\pi/2$.
We thus see that the term proportional to $b_\pm$ in
$F^{(5)}_{rt\theta\phi y}$ is singular for $\theta=\pi/2$
and the term proportional to $d_\pm$ in $F^{(5)}_{ry\theta t\psi}$ is
singular at $\theta=0$. Therefore we have to
take $b_\pm=d_\pm=0$.

Consider now the behaviour around $f=0$ (i.e. $r=0$ and $\theta=\pi/2$),
where the metric goes to
\bea
{ds^2\over \sqrt{Q_1 Q_2}}&\approx& {dr^2\over r^2+a^2}+{r^2\over Q_1
Q_2}\,dy^2-{r^2+a^2\over Q_1 Q_2}\,
\Bigl(1-2{a^2\over\sqrt{Q_1 Q_2}})
+d\theta^2+\cos^2\theta\,d\psi'^2\\&+&
\sin^2\theta\,d\phi'^2\Bigl(1+2{a^2\over\sqrt{Q_1 Q_2}}\Bigr)+
4\,{a\, r^2\over\sqrt{Q_1 Q_2}}\,\cos^2\theta\,dy
\,d\psi'+4\,{a\,(r^2+a^2)\over\sqrt{Q_1 Q_2}}\,
\sin^2\theta\,dt\, d\phi'\nonumber
\eea
Then we have
\bea
&&F^{(5)}_{rt\theta\phi'\psi'}\approx -a_-\,Q_1
Q_2\,{\sin\theta\cos\theta\over r}\nonumber\\
&&F^{(5)}_{rt\theta\phi' y}\approx -2
a_-\,a\,r\,\sin\theta\,\cos\theta\nonumber\\
&&F^{(5)}_{ry\theta\phi'\psi'}\approx c_-\,Q_1 Q_2\,\Bigl(1+2{a^2\over
\sqrt{Q_1 Q_2}}\Bigr)
\,{\sin\theta\cos\theta\,r\over r^2+a^2}\nonumber\\
&&F^{(5)}_{ry\theta t\psi'}\approx 2 c_-\,a\,r\,\sin\theta\,\cos\theta
\eea
Regularity at $f=0$ then requires $a_-=0$ (and thus $c_+=-\mu_2$), while
$c_-$ is left arbitrary.

Let us now consider the behaviour of the field strength at asymptotic
infinity:
\bea
F^{(5)}&=&a_+\,dr\wedge dt\wedge d\theta\wedge d\phi'\wedge
{g_{\psi'\psi'}\,d\psi'+
g_{\psi'y}\,dy\over
\sqrt{-g}\,g^{rr}\,g^{\theta\theta}\,(g^{tt}g^{\phi\phi}-
g^{t\phi}g^{t\phi})}\nonumber\\
&+&c_+\,dr\wedge dy\wedge d\theta\wedge {g_{\phi'\phi'}\,d\phi'+g_{\phi'
t}\,dt \over
\sqrt{-g}\,g^{rr}\,g^{\theta\theta}\,(g^{yy}g^{\psi\psi}-
g^{y\psi}g^{y\psi})}\wedge
d\psi'\nonumber\\
&\approx& -a_+\,r^3\,\sin\theta\,\cos\theta\,dr\wedge dt\wedge
d\theta\wedge d\phi\wedge d\psi\nonumber\\
&+& c_+\,r^3\,\sin\theta\,\cos\theta\,dr\wedge dy\wedge d\theta\wedge
d\phi\wedge d\psi
\nonumber\\
\eea
The formula above shows that, asymptotically, the field strength is
constant in local orthonomal coordinates.

We have the freedom to choose the constant $c_-$ to have any value that
we want; this freedom corresponds to adding a
  smooth magnetic field everywhere to the background. A simple choice of
$c_-$ would be the one that makes $c_+=0$, so that
this magnetic field vanishes at infinity. Then we get
\bea
F^{(5)}&=&-\mu_2\,dr\wedge dt\wedge d\theta \wedge d\phi'\wedge
{g_{\psi'\psi'}\,d\psi'+g_{\psi'y}\,dy\over
\sqrt{-g}\,g^{rr}\,g^{\theta\theta}\,(g^{tt}g^{\phi'\phi'}-
g^{t\phi'}g^{t\phi'})}\nonumber\\
&=&\mu_2\,{\sin\theta\,\cos\theta\over r}\,dr\wedge dt\wedge d\theta
\wedge d\phi'\wedge
\Bigl[h^2\,f\Bigl(r^2 +{Q_1 Q_2\,a^2\,\cos^2\theta\over h^2
f^2}\Bigr)d\psi'\nonumber\\
&&\qquad\qquad\qquad+{a\over\sqrt{Q_1 Q_2}}\,r^2\,(f+Q_1+Q_2)\,dy\Bigl]
\,,\,{\rm for}\,r>r_0\nonumber\\
F^{(5)}&=&-\mu_2\,dr\wedge dy\wedge d\theta\wedge d\psi'\wedge
{g_{\phi'\phi'}\,d\phi'+g_{\phi't}\,dt\over
\sqrt{-g}\,g^{rr}\,g^{\theta\theta}\,(g^{tt}g^{\psi'\psi'}-
g^{y\psi'}g^{y\psi'})}\nonumber\\
&=&-\mu_2\,{r\,\sin\theta\,\cos\theta\over r^2+a^2}\,dr\wedge dy\wedge
d\theta\wedge d\psi'\wedge
\Bigl[h^2\,f\Bigl(r^2+a^2 -{Q_1 Q_2\,a^2\,\sin^2\theta\over h^2
f^2}\Bigr)d\phi'\nonumber\\
&&\qquad\qquad\qquad+{a\over\sqrt{Q_1 Q_2}}
\,(r^2+a^2)\,(f+Q_1+Q_2)\,dt\Bigl]\,,\,{\rm for}\,r<r_0
\eea

For any choice of $c_-$ we find the the stress tensor of the field goes
to a constant rather than vanish at infinity.
We can thus generate a uniform cosmological type contribution in the
spacetime dimensionally reduced on the $y$ direction. The
only way to cancel this contribution would be to have a $\overline
{D3}$ brane in the `throat' of the microstate geometry, or in
the throat of a different microstate geometry located at some other
spacetime point. We
have to be aware that the energy computed from the DBI action in the
above sections does not iclude this (possibly divergent) field
contribution.

\section{Supersymmetry properties of the branes}\label{susy}

The simple expressions for the energies as a sum of the contribution
from individual charges signifies a threshold bound state. As
is usual in such situations, this usually follows from supersymmetry
and BPS bounds. In this section we will examine the supersymmetry
properties of these brane configurations.

\subsection{Supersymmetry of the $D2$ brane}\label{susyd2}

In this section we will examine the supersymmetry
properties for the case of $D2$ branes in IIA theory. The
considerations can be easily generalized to the M-branes.

\subsubsection{Killing spinors of the near-horizon background}
 We work in global coordinates.
The metric, dilaton, RR 1-form, and RR 3-forms of the near-horizon
background were given in (\ref{nefifteen}).
We use $m,n...=\tau,\chi,\theta,\phi,1,...,6$ as the ten-dimensional
curved space indices,
$a,b...=\hat{\tau},\hat{\chi},\hat{\theta},\hat{\phi},\hat{1},...,\hat{6}$
(or sometimes, equivalently,
$a,b...=\underline{0},...,\underline{9}$) as the tangent space
indices. The Clifford algebra is \BE \left\{\Gamma^a,\Gamma^b
\right\}=2\eta^{ab}\EE with $\eta^{ab}$ having signature
$(-,+,...,+)$, and the gamma matrices $\Gamma^a$'s are $32$ by $32$
real matrices ($\Gamma^{\hat{\tau}}$ being antisymmetric, and
$\Gamma^{\hat{\chi}},...,\Gamma^{\hat{6}}$ being symmetric).
$\Gamma^{\underline{10}}\equiv\Gamma^{\underline{0}...\underline{9}}$
and $\left(\Gamma^{\underline{10}}\right)^2=1$. We use
$32$-component real spinors $y$, and define $\bar{y}\equiv
y^T\Gamma^{\underline{0}}$.

The local supersymmetry variation of the dilatino, parameterized by
a $32$-component real spinor $\epsilon$, is \BE
\delta\lambda=\frac{1}{8}e^\Phi\left(\frac{3}{2!}F^{(2)}_{ab}\Gamma^{ab}\Gamma^{\underline{
\varphi}} +\frac{1}{4!}F^{(4)}_{abcd}\Gamma^{abcd}\right)\epsilon\EE
and the gravitino variation is \BE
\delta\psi_m=\left[\partial_m+\frac{1}{4}\omega_{mab}\Gamma^{ab}
+\frac{1}{8}e^\Phi\left(\frac{1}{2!}F^{(2)}_{ab}\Gamma^{ab}\Gamma_m\Gamma^{\underline{\varphi}}
+\frac{1}{4!}F^{(4)}_{abcd}\Gamma^{abcd}\Gamma_m
\right)\right]\epsilon\EE where $\Gamma^{\underline{\varphi}}\equiv
-\Gamma^{\underline{10}}=-\Gamma^{\hat{\tau}\hat{\chi}\hat{\theta}\hat{\phi}\hat{1}
\hat{2}\hat{3}\hat{4}\hat{5}\hat{6}}$. Plugging in the expressions
of the RR field strength, we get \BE
\delta\lambda=\frac{1}{R}N\epsilon, \ \
\delta\psi_m=\left[\partial_m+\frac{1}{4}\omega_{mab}\Gamma^{ab}+\frac{1}{R}M\Gamma_m\right]\epsilon
\EE where the matrices $N$ and $M$ are given by \BE &
&N=\frac{1}{8}\left[3\Gamma^{\hat{\theta}\hat{\phi}\hat{1}
\hat{2}\hat{3}\hat{4}\hat{5}\hat{6}}+\Gamma^{\hat{\theta}\hat{\phi}}\left(
\Gamma^{\hat{1}\hat{2}}+\Gamma^{\hat{3}\hat{4}}+\Gamma^{\hat{5}\hat{6}}\right)\right]\nonumber\\
& &M=\frac{1}{8}\left[-\Gamma^{\hat{\theta}\hat{\phi}\hat{1}
\hat{2}\hat{3}\hat{4}\hat{5}\hat{6}}+\Gamma^{\hat{\theta}\hat{\phi}}\left(
\Gamma^{\hat{1}\hat{2}}+\Gamma^{\hat{3}\hat{4}}+\Gamma^{\hat{5}\hat{6}}\right)\right]
\EE (note the only nonvanishing
$\frac{1}{4}\omega_{mab}\Gamma^{ab}$'s are $\frac{1}{4}\omega_{\tau
ab}\Gamma^{ab}=-\frac{\sinh\chi}{2}\Gamma^{\hat{\tau}\hat{\chi}}$
and $\frac{1}{4}\omega_{\phi
ab}\Gamma^{ab}=\frac{\cos\theta}{2}\Gamma^{\hat{\phi}\hat{\theta}}$;
also note that $\Gamma^{\hat{\theta}\hat{\phi}\hat{1}
\hat{2}\hat{3}\hat{4}\hat{5}\hat{6}}=-\left(\Gamma^{\hat{\theta}\hat{\phi}\hat{1}
\hat{2}}\right)\left(\Gamma^{\hat{\theta}\hat{\phi}\hat{3}
\hat{4}}\right)\left(\Gamma^{\hat{\theta}\hat{\phi}\hat{5}
\hat{6}}\right)$.) Next we solve $\delta\lambda=0$ and
$\delta\psi_m=0$ to find the Killing spinors.

Let's divide the $32$-dimensional vector space of $\epsilon$ into
eight subspaces of simultaneous eigenvectors of
$\Gamma^{\hat{\theta}\hat{\phi}\hat{1} \hat{2}}$,
$\Gamma^{\hat{\theta}\hat{\phi}\hat{3} \hat{4}}$, and
$\Gamma^{\hat{\theta}\hat{\phi}\hat{5} \hat{6}}$, labeled as
$(\pm\pm\pm)$ (with the $\pm$'s denotes the $\pm 1$ eigenvalues of
these three matrices, respectively). Each of these subspaces is
four-dimensional by itself. It is easy to see that,
$\delta\lambda=0$ if and only if \BE \epsilon=\epsilon_+
+\epsilon_-\EE with $\epsilon_+\in (+++)$ and $\epsilon_-\in (---)$.

Plugging the above expression for $\epsilon$ into $\delta\psi_m$ and
integrating, we then get the explicit expression of the eight
Killing spinors of $AdS_2\times S^2\times T^6$, four of them being
\BE\label{killingspinor1}\epsilon_1=\left[e^{-\frac{1}{2}\chi\Gamma^{\hat{\chi}}}
e^{\frac{1}{2}\tau\Gamma^{\hat{\tau}}}\sin\frac{\theta}{2}
e^{\frac{1}{2}\phi\Gamma^{\hat{\phi}\hat{\theta}}} +
e^{\frac{1}{2}\chi\Gamma^{\hat{\chi}}}
e^{-\frac{1}{2}\tau\Gamma^{\hat{\tau}}}\left(-\cos\frac{\theta}{2}\right)\Gamma^{\hat{\theta}}
e^{\frac{1}{2}\phi\Gamma^{\hat{\phi}\hat{\theta}}}\right]\Phi_0\EE
with $\Phi_0$ being an arbitrary constant $32$-component real spinor
in the four-dimensional $(+++)$ subspace, i.e.
$\Gamma^{\hat{\theta}\hat{\phi}\hat{1} \hat{2}}\Phi_0=\Phi_0$,
$\Gamma^{\hat{\theta}\hat{\phi}\hat{3} \hat{4}}\Phi_0=\Phi_0$, and
$\Gamma^{\hat{\theta}\hat{\phi}\hat{5} \hat{6}}\Phi_0=\Phi_0$; and
the other four being \BE\label{killingspinor2} \epsilon_2=
\left[e^{-\frac{1}{2}\chi\Gamma^{\hat{\chi}}}
e^{\frac{1}{2}\tau\Gamma^{\hat{\tau}}}\cos\frac{\theta}{2}
e^{-\frac{1}{2}\phi\Gamma^{\hat{\phi}\hat{\theta}}} +
e^{\frac{1}{2}\chi\Gamma^{\hat{\chi}}}
e^{-\frac{1}{2}\tau\Gamma^{\hat{\tau}}}\sin\frac{\theta}{2}\Gamma^{\hat{\theta}}
e^{-\frac{1}{2}\phi\Gamma^{\hat{\phi}\hat{\theta}}}\right]\Phi_0'\EE
with $\Phi_0'$ being another arbitrary constant $32$-component real
spinor in the four-dimensional $(+++)$ subspace. A general Killing
spinor is given by $\epsilon=\epsilon_1+\epsilon_2$.

\subsubsection{Supersymmetric D2 configuration}\label{subsubsection D2 susy config} Next we show that the D2 trajectory considered in
Subsection \ref{IIABH} preserves half of the
background supersymmetries. Recall that the trajectory is \BE
\tau=\chi^0,\ \theta=\chi^1,\ \phi=\chi^2,\ \chi=\chi_0, \
y_1=0,...,y_6=0\EE for which the $\kappa$ projection matrix as given
in \cite{Bergshoeff:1996tu} evaluates to
 \BE\label{eqn kappa projection matrix}
\Gamma=\frac{-1}{\cosh\chi_0}\left(1+\sinh\chi_0\Gamma^{\hat{\theta}\hat{\phi}}\Gamma^{\underline{10}}\right)
\Gamma^{\hat{\tau}\hat{\theta}\hat{\phi}}\EE The supersymmetries
preserved by the D2 brane are the Killing spinors $\epsilon$ that
satisfy \BE (1-\Gamma)\epsilon=0\EE After some manipulation, one
finds that there are four supersymmetries preserved, with two of the
corresponding Killing spinors given by eqn. (\ref{killingspinor1})
constrained by
$(1+\Gamma^{\hat{\tau}\hat{\theta}\hat{\phi}})\Phi_0=0$, and the
other two given by eqn. (\ref{killingspinor2}) constrained by
$(1+\Gamma^{\hat{\tau}\hat{\theta}\hat{\phi}})\Phi_0'=0$. Note that
these projection conditions turn out to be independent of the D0
charge (i.e., independent of the value of $\chi_0$).

\subsubsection{Topological charge of the brane}\label{subsubsection central charge of
D2} In \cite{Hackett-Jones:2003vz} $p$-forms constructed from
background Killing spinors are integrated over probe branes' spatial
worldvolumes to give topological charges in M-theory.
\cite{Saffin:2004ar} generalize this to IIA theory, whose approach
we shall now apply to the above D2 brane. We shall find a central
charge $C_{D2}=M_2e^{-\Phi}+M_0e^{-\Phi}$ in the superalgebra, which
equals the D2's global energy and shows that the D2 indeed saturates
a BPS bound.

After being sandwiched between $\epsilon^T$ and $\epsilon$ (where
$\epsilon$ is a Killing spinor, and is treated as a commuting rather
than anti-commuting variable), the superalgebra with the probe brane
can be written as \BE\label{eqn D2susyalgebra}
\left(Q\epsilon\right)^2=\int d^2\chi K_\mu p^\mu\pm \int
\omega_{D2}\EE where the integrals are over the spatial worldvolume
of the brane, $K$ is a one-form defined as a bilinear of $\epsilon$
\BE K=\bar{\epsilon}\Gamma_a\epsilon\ e^a\EE ($e^a$ being the
vielbein one-form) and $\omega_{D2}$ is a closed two-form also
constructed from bilinears of $\epsilon$. The choice of
$\omega_{D2}$ is background-specific\footnote{For a string probe,
there is a general expression for the closed one-form
$\omega_{string}$, see \cite{Saffin:2004ar} for details.}, and we
shall take the one used in \cite{Saffin:2004ar} to consider
supertubes \BE \omega_{D2}=\mu_2\left(e^{-\Phi}\Omega+K\cdot
A^{(3)}+\tilde{K}\wedge A^{(1)}- 2\pi\alpha'F\right)\EE with the
$\cdot$ denoting the inner product of $q$-forms with $p$-forms
($p<q$)
$\left(\alpha_p\cdot\beta_q\right)_{a_1....a_{q-p}}=\left(1/p!\right)
\alpha^{b_1...b_p}\beta_{b_1...b_pa_1...a_{q-p}}$ , and \BE
\Omega=\frac{1}{2}\bar{\epsilon}\Gamma_{ab}\epsilon\ e^{ab},\ \
\tilde{K}=\bar{\epsilon}\Gamma_a\Gamma^{\underline{10}}\epsilon\
e^a\EE Note that our choice for the contribution of the worldvolume
field strength to $\omega_{D2}$ differs from that of
\cite{Saffin:2004ar} by a minus sign. Due to $\epsilon$'s being a
Killing spinor, $K$ turns out to be a Killing vector, and $K,
\Omega, \tilde{K}$ satisfy the following differential relations
(which are obtained by plugging our background into equations (3.18)
and (3.19) of \cite{Saffin:2004ar}) \BE d\tilde{K}=0,\ \
d\left(e^{-\Phi}\Omega\right)=\tilde{K}\wedge F^{(2)}+K\cdot
F^{(4)}\EE Using these relations one finds \BE d\omega_{D2}=K\cdot
F^{(4)}+d\left(K\cdot A^{(3)}\right)={\cal L}_K A^{(3)}\EE Hence
$\omega_{D2}$ will be closed if $A^{(3)}$ is invariant under the Lie
derivative ${\cal L}_K$, and now we turn our attention to $K$.

One readily sees that $K_{\hat{1}}=0$, since $\epsilon$ only has
components in $(+++)$ and $(---)$ while $\Gamma^{\tau\hat{1}}$ takes
$(+++)$ to $(-++)$ and $(---)$ to $(+--)$, and orthogonality of the
subspaces then gives
$\epsilon^T\Gamma^{\hat{\tau}\hat{1}}\epsilon=0$. Similiarly,
$K_{\hat{2}},...,K_{\hat{6}}$ all vanish.

After some algebra, one finds \BE
K_{\hat{\chi}}=\epsilon^T\Gamma^{\hat{\tau}\hat{\chi}}\epsilon=\cos\tau\left(\Phi_0^T\Gamma^{\hat{\tau}\hat{\chi}}
\Phi_0+{\Phi'}_0^T\Gamma^{\hat{\tau}\hat{\chi}}\Phi'_0\right)
+\sin\tau\left(\Phi_0^T\Gamma^{\hat{\chi}}
\Phi_0+{\Phi'}_0^T\Gamma^{\hat{\chi}}\Phi'_0\right) \EE \BE
K_{\hat{\theta}}=\epsilon^T\Gamma^{\hat{\tau}\hat{\theta}}\epsilon=2\Phi_0^T\Gamma^{\hat{\tau}}
\exp\left(-\phi\Gamma^{\hat{\phi}\hat{\theta}}\right)\Phi'_0\EE \BE
K_{\hat{\phi}}=\epsilon^T\Gamma^{\hat{\tau}\hat{\phi}}\epsilon=2\cos\theta{\Phi'}_0^T
\Gamma^{\hat{\tau}\hat{\theta}\hat{\phi}}\exp\left(\phi\Gamma^{\hat{\phi}\hat{\theta}}\right)\Phi_0
+\sin\theta\left(\Phi_0^T\Gamma^{\hat{\tau}\hat{\theta}\hat{\phi}}\Phi_0
-{\Phi'}_0^T\Gamma^{\hat{\tau}\hat{\theta}\hat{\phi}}\Phi'_0\right)\EE

Now let's pick out a unique Killing spinor by further imposing the
projection and normalization conditions \BE &
&\Gamma^{\hat{\tau}\hat{\theta}\hat{\phi}}\Phi_0=-\Phi_0,\ \
\Gamma^{\hat{\tau}\hat{1}\hat{2}\hat{3}\hat{4}\hat{5}\hat{6}}\Phi_0=\Phi_0\nonumber\\
& &\Gamma^{\hat{\tau}\hat{\theta}\hat{\phi}}\Phi'_0=-\Phi'_0,\ \
\Gamma^{\hat{\tau}\hat{1}\hat{2}\hat{3}\hat{4}\hat{5}\hat{6}}\Phi'_0=-\Phi'_0\nonumber\\
& &\Phi_0^T\Phi_0=\frac{\Delta}{2},\ \
{\Phi'}_0^T\Phi'_0=\frac{\Delta}{2}\EE where $\Delta$ is some
positive normalization number whose value shall be determined soon.
Note that this Killing spinor is preserved by the D2 (see subsection
\ref{subsubsection D2 susy config}). For this Killing spinor, one
immediately finds \BE K_{\hat{\chi}}=0,\ K_{\hat{\theta}}=0,\
K_{\hat{\phi}}=0, and, \
K_{\hat{\tau}}=\epsilon^T\epsilon=\Delta\cosh\chi\EE i.e.
$K=\frac{-\Delta}{R}\frac{\partial}{\partial\tau}$, which is the
Killing vector generating global time translation. For this $K$
${\cal L}_K A^{(3)}$ vanishes, and we then find $\omega_{D2}$ is
indeed closed. (Actually, the story here is quite trivial: since
$A^{(3)}, F^{(4)}$ don't have any $\tau$-component, $K\cdot A^{(3)},
K\cdot F^{(4)}$ both vanish. )

Having established the closedness of $\omega_{D2}$, we now integrate
it over the spatial worldvolume of the D2. Since $A^{(1)}\sim d\tau$
and $K\cdot A^{(3)}$ vanishes, only the $e^{-\Phi}\Omega$ term and
the worldvolume flux term contributes to the integral \BE
\int_{S^2}\omega_{D2}&=&
\mu_2\int_{S^2}\frac{R}{q_0}\left(\epsilon^T\Gamma^{\hat{\tau}\hat{\theta}\hat{\phi}}\epsilon\right)
R^2\sin\theta d\theta\wedge
d\phi-\mu_22\pi\alpha'\int_{S^2}F\nonumber\\
&=&-4\pi\mu_2\Delta\frac{R^3}{q_0}-M_0=-\Delta M_2e^{-\Phi}-M_0\EE
where in the second line we've used the fact that
$\epsilon^T\Gamma^{\hat{\tau}\hat{\theta}\hat{\phi}}\epsilon$
evaluates to $-\Delta$ for the particular Killing spinor we've
chosen. Since $\int d^2\chi K_\mu p^\mu=K^\tau
P_\tau=-\frac{\Delta}{R}P_\tau=-\Delta E$ (recall that the physical
energy is $E=P_\tau/R$), and the particular $\epsilon$ is perserved
by the D2, the supersymmetry algegra (\ref{eqn D2susyalgebra})
becomes \BE -\Delta E=\mp\left(-\Delta M_2e^{-\Phi}-M_0\right)\EE
which upon taking the lower sign on the r.h.s. gives \BE
E=C_{D2}=M_2e^{-\Phi}+\frac{M_0}{\Delta}\EE From this we see that we
should take the normalization number $\Delta$ to be
$e^{\Phi}=\frac{q_0}{R}$, which results in
$C_{D2}=M_2e^{-\Phi}+M_0e^{-\Phi}$. This is the same as the global
energy we computed earlier for this D2 trajectory and shows this D2
saturates the BPS bound.

\subsubsection{Supersymmetry of D2 in the full black hole
geometry}\label{subsubsection susy of D2 in full geometry} In this
subsection, we show that the D2 considered above does not preserve
any of the supersymmetries of the full black hole geometry (except
in the $\chi_0\rightarrow\infty$ limit where it is effectively a
bunch of D0 branes), and is thus not really a stable configuration
in the full geometry. First let's work out the Killing spinors of
the full geometry.

Recall that the metric of the full geometry is given by \BE ds^2=&
&\frac{-1}{\sqrt{H_0H_1H_2H_3}}dt^2+\sqrt{H_0H_1H_2H_3}\left(dr^2+r^2d\Omega_2^2\right)
+\sqrt{\frac{H_0H_1}{H_2H_3}}\left(dy_1^2+dy_2^2\right)\nonumber\\
& &+\sqrt{\frac{H_0H_2}{H_1H_3}}\left(dy_3^2+dy_4^2\right)
+\sqrt{\frac{H_0H_3}{H_1H_2}}\left(dy_5^2+dy_6^2\right)\EE where
$H_0=1+\frac{q_0}{r}$, $H_i=1+\frac{p_i}{r}, i=1,2,3$. The
nonvanishing components of the RR four-form and two-form field
strengths are given by \BE& & F^{(4)}_{\theta\phi
12}=-\frac{dH_1}{dr}r^2\sin\theta,\
 F^{(4)}_{\theta\phi 34}=-\frac{dH_2}{dr}r^2\sin\theta,\ \ F^{(4)}_{\theta\phi 56}=-\frac{dH_3}{dr}r^2\sin\theta
 \nonumber\\
 & &F^{(2)}_{rt}=-\frac{1}{(H_0)^2}\frac{dH_0}{dr}\EE
And the dilaton is \BE
e^\Phi=\left(\frac{H_1H_2H_3}{(H_0)^3}\right)^{-1/4}\EE Note that
the dilaton is no longer constant once we go beyond the near-horizon
region. Now the local supersymmetry variation of the dilatino is
given by \BE \delta\lambda=\left[\frac{1}{2}\Gamma^m\partial_m\Phi
+\frac{1}{8}e^\Phi\left(\frac{3}{2!}F^{(2)}_{ab}\Gamma^{ab}\Gamma^{\underline{
\varphi}}
+\frac{1}{4!}F^{(4)}_{abcd}\Gamma^{abcd}\right)\right]\epsilon\EE
which after plugging in the expression of the RR fields becomes \BE
\delta\lambda=\frac{1}{8}\left(H_0H_1H_2H_3\right)^{-1/4}\hspace{-0.2in}&
&
\Bigg\{-\left(\sum_{i=1}^3\frac{1}{H_i}\frac{dH_i}{dr}-\frac{3}{H_0}\frac{dH_0}{dr}
\right)\Gamma^{\hat{r}}+\Bigg[\frac{-3}{H_0}\frac{dH_0}{dr}\Gamma^{\hat{\theta}\hat{\phi}\hat{1}\hat{2}
\hat{3}\hat{4}\hat{5}\hat{6}}\nonumber\\
&
&+\left(-\frac{1}{H_1}\frac{dH_1}{dr}\Gamma^{\hat{\theta}\hat{\phi}\hat{1}\hat{2}}
-\frac{1}{H_2}\frac{dH_2}{dr}\Gamma^{\hat{\theta}\hat{\phi}\hat{3}\hat{4}}
-\frac{1}{H_3}\frac{dH_3}{dr}\Gamma^{\hat{\theta}\hat{\phi}\hat{5}\hat{6}}\right)
\Bigg] \Bigg\}\epsilon\EE Now we divide the $32$-component spinor
$\epsilon$ into sixteen subspaces labeled by $(s_1s_2s_3w)$ with
$s_1,s_2,s_3=\pm 1$ being eigenvalues of
$\Gamma^{\hat{\theta}\hat{\phi}\hat{1}\hat{2}}$,
$\Gamma^{\hat{\theta}\hat{\phi}\hat{3}\hat{4}}$,
$\Gamma^{\hat{\theta}\hat{\phi}\hat{5}\hat{6}}$, and $w=\pm 1$ being
eigenvalue of $\Gamma^{\hat{r}}$. It is then easy to see that,
$\delta\lambda=0$ if and only if \BE
\epsilon=\epsilon_{+++-}+\epsilon_{---+}\EE where the subscripts
denote the subspace the spinors belong to. This gives us the four
Killing spinors of the full black hole geometry, and we shall denote
them as $\epsilon_{full}$. The concrete coordinate-dependence of
$\epsilon_{full}$ can be worked out by requiring the vanishing of
the gravitino variation, however we don't need this detailed
knowledge for the analysis below.

Now let's look at the kappa-projection matrix $\Gamma$ given in eqn.
(\ref{eqn kappa projection matrix}) in the near-horizon region. Note
that \BE\label{eqn useful matrix
identity}\Gamma^{\hat{\theta}\hat{\phi}}\Gamma^{\underline{10}}\Gamma^{\hat{\tau}\hat{\theta}\hat{\phi}}
=\Gamma^{\hat{\chi}}\Gamma^{\hat{\theta}\hat{\phi}\hat{1}\hat{2}}\Gamma^{\hat{\theta}\hat{\phi}\hat{3}\hat{4}}
\Gamma^{\hat{\theta}\hat{\phi}\hat{5}\hat{6}}\EE and that
$\Gamma^{\hat{\chi}}$ is the same as $\Gamma^{\hat{r}}$ because both
are tangent-indiced gamma matrices. We see that $\Gamma$ commutes
with $\Gamma^{\hat{\theta}\hat{\phi}\hat{1}\hat{2}}$,
$\Gamma^{\hat{\theta}\hat{\phi}\hat{3}\hat{4}}$,
$\Gamma^{\hat{\theta}\hat{\phi}\hat{5}\hat{6}}$. Hence requiring the
supersymmetry of the full geometry to be preserved by D2, i.e., \BE
\Gamma\epsilon_{full}=\epsilon_{full} \EE is equivalent to requiring
\BE\Gamma\epsilon_{+++-}=\epsilon_{+++-},\ and\ \
\Gamma\epsilon_{---+}=\epsilon_{---+} \EE which is immediately seen
to be impossible to satisfy for any finite value of $\chi_0$,
because \BE\label{eqn gamma epsilonfull}& &
\Gamma\epsilon_{+++-}=\frac{-1}{\cosh\chi_0}\Gamma^{\hat{\tau}\hat{\theta}\hat{\phi}}
\epsilon_{+++-}+\tanh\chi_0\ \epsilon_{+++-}\nonumber\\
&
&\Gamma\epsilon_{---+}=\frac{-1}{\cosh\chi_0}\Gamma^{\hat{\tau}\hat{\theta}\hat{\phi}}
\epsilon_{---+}+\tanh\chi_0\ \epsilon_{---+}\EE (where the identity
(\ref{eqn useful matrix identity}) has been used) and we see that
the first terms on the right hand sides have the wrong eigenvalue
under $\Gamma^{\hat{r}}$ (because
$\Gamma^{\hat{\tau}\hat{\theta}\hat{\phi}}$ anticommutes with
$\Gamma^{\hat{r}}$). This proves our claim that, for finite $\chi_0$
the D2 brane doesn't preserve any of the four usual supersymmetries
of the full geometry (the four supersymmetries preserved by the D2
as shown in subsection \ref{subsubsection D2 susy config} have to be
formed out of linear combinations of the usual supersymmetries of
the full geometry and the conformal supersymmetries that are present
only in the near-horizon region). What about the case
$\chi_0\rightarrow\infty$? In this case, the first terms on the
right hand sides of eqn. (\ref{eqn gamma epsilonfull}) vanish, and
the second terms become $\epsilon_{+++-}$ and $\epsilon_{---+}$
respectively, giving exactly what is needed for
$\Gamma\epsilon_{full}=\epsilon_{full}$. This comes as no surprise
since in the infinite $\chi_0$ limit the D2 has an infinite D0
charge and is effectively just a bunch of D0 branes, which is known
to preserve all the four usual supersymmetries of the full black
hole geometry.

\subsection{Supersymmetry of D3 branes in Microstate
geometry}\label{susymm}

In this subsection we examine supersymmetry properties of D3 brane
in the 2 charge microstate geometry discussed in section
(\ref{microstate}). Analogously to the D2 case considered above, we
shall find that the D3 brane preserves half of the supersymmetries
of the near-hoziron geometry, but doesn't preserve any of the
supersymmetries of the full asymptotically flat geometry.

As in the above IIA case, we use
$m,n...=t,y,r,\theta,\phi,\psi,y^1,y^2,y^3,y^4$ to denote curved
space indices, and $a,b...=\hat{0},\hat{1},...,\hat{9}$ to denote
tangent space indices. $\hat{\Gamma}_a$ are ten dimensional Gamma
matrices, which we will decompose into direct products of  $6$-d
Gamma matrices denoted as $\tilde{\Gamma}_{a}$ and $4$-d Gamma
matrices denoted as $\Gamma_a$. The analysis in the near-horizon
region $AdS_3\times S^3\times T^4$ is similar to the D2 case, hence
instead of giving all the details here we will simply quote the
near-horizon results when needed without proof.

Let us consider the D3 brane at its stable point $\dot{y}=-1$. (In
Section \ref{microstate} the choice of $\dot{y}=+1$ was made. This
difference in choices does not affect the conclusion of the analysis
below, because they just correspond to conjugate Killing spinors
preserved by the D3 brane). As shown in Section \ref{microstate},
for $\dot{y}=-1$ the determinant of the metric induced on the brane
simplifies to $\sqrt{-g}= a\sqrt{Q_{1}Q_{2}}\sin\theta\cos\theta$.
Then the kappa symmetry condition (after getting rid of
antisymmetrization and combinatorial factors) becomes \be
\gamma_{t}\gamma_{\theta}\gamma_{\psi}\gamma_{\phi}\xi =
-ia\sqrt{Q_{1}Q_{2}}\sin\theta\cos\theta\xi \ee where $\gamma_i$ are
the pull backs on the brane worldvolume of the space time Gamma
matrices. Using the vielbeins for the six dimensional 2-charge
microstate metric
\begin{eqnarray}
e^{\hat{0}}= \frac{1}{\sqrt{h}}\Bigl(dt +
\frac{a\sqrt{Q_{1}Q_{2}}}{f}\sin^{2}\theta d\phi\Bigr) \  ,  \
e^{\hat{1}}= \frac{1}{\sqrt{h}}
\Bigl(dy - \frac{a\sqrt{Q_{1}Q_{2}}}{f}\cos^{2}\theta d\psi\Bigr) \ \  \\
e^{\hat{2}} = \sqrt{\frac{hf}{r^{2}+a^{2}}}dr \ ,\
e^{\hat{3}}=\sqrt{hf}d\theta \ , \ e^{\hat{4}} = \sqrt{h}r\cos\theta
d\psi \ , \ e^{\hat{5}}= \sqrt{h(r^{2}+a^{2})}\sin\theta d\phi \ \
\end{eqnarray}
the induced gamma matices are found to be
\begin{eqnarray}
\gamma_{\theta}= e^{\hat{3}}_{\theta}\tilde{\Gamma}_{\hat{3}} \ , \
\gamma_{\phi} = e^{\hat{5}}_{\phi}\tilde{\Gamma}_{\hat{5}} +
e^{\hat{0}}_{\phi} \tilde{\Gamma}_{\hat{0}} \ , \ \gamma_{t}=
e^{\hat{0}}_{t}\tilde{\Gamma}_{\hat{0}} +
\dot{y}e^{\hat{1}}_{y}\tilde{\Gamma}_{\hat{1}} \ , \ \gamma_{\psi}=
e^{\hat{4}}_{\psi}\tilde {\Gamma}_{\hat{4}} +
e^{\hat{1}}_{\psi}\tilde{\Gamma}_{\hat{1}}
\end{eqnarray}
Setting $\dot{y}=-1$ in the expression for $\gamma_t$ and using
using $e^{\hat{0}}_{t}=e^{\hat{1}}_{y}$ we can then rewrite the
kappa symmetry matrix in terms of constant Gamma matrices and
vielbeins as \be \gamma_{t\theta\psi\phi}\,\xi =
e^{\hat{0}}_{t}e^{\hat{3}}_{\theta}\left[
(e^{\hat{1}}_{\psi}e^{\hat{5}}_{\phi}\tilde{\Gamma}_{\hat{3}}
\tilde{\Gamma}_{\hat{5}}-
e^{\hat{4}}_{\psi}e^{\hat{0}}_{\phi}\tilde{\Gamma}_{\hat{3}}\tilde{\Gamma}_{\hat{4}})(1-
\tilde{\Gamma}_{\hat{0}\hat{1}}) +(e^{\hat{4}}_{\psi}
e^{\hat{5}}_{\phi}\tilde{\Gamma}_{\hat{0}}\tilde{\Gamma}_{\hat{3}}\tilde{\Gamma}_{\hat{4}}\tilde
{\Gamma}_{\hat{5}} - e^{\hat{1}}_{\psi}e^{\hat{0}}_{\phi}
\tilde{\Gamma}_{\hat{0}}\tilde{\Gamma}_{\hat{3}})(1+\tilde{\Gamma}_{\hat{0}\hat{1}})
\right]\xi \ee

Now let's look at a Killing spinor for the asymptotically flat
metric generated by the background $D3$ branes. We know that it will
be of the form $\xi = g(x)\xi_{0}$ (see, e.g.,
\cite{Emparan:2001ux}). Here $g(x)$ is some spacetime dependent part
which will cancel from both sides of kappa symmetry matrix as it
doesn't depend on gamma matrices (unlike the near horizon case). The
constant part $\xi_{0}$ satisfies projection conditions
corresponding to two orthogonal sets of $D3$ branes. Our $D3$ branes
are along directions $y67$ and $y89$, hence
\begin{eqnarray}
\xi_{0} + i\hat{\Gamma}_{\hat{0}\hat{1}\hat{6}\hat{7}}\xi_{0} =0 \ ,
\ \xi_{0} + i\hat{\Gamma}_{\hat{0}\hat{1}\hat{8}\hat{9}}\xi_{0} =0
\end{eqnarray}

We decompose the constant spinor $\xi_{0}$ as $\xi^{(0)}_{M6}\otimes
\xi^{(0)}_{T4}$. This gives three constraints
\begin{eqnarray}
\xi^{(0)}_{M6} + \tilde{\Gamma}_{\hat{0}\hat{1}}\xi^{(0)}_{M6} =0 \
, \ \xi^{(0)}_{T4} + i\Gamma_{\hat{6}\hat{7}}\xi^{(0)}_{T4} =0 \ , \
\xi^{(0)}_{T4} + i\Gamma_{\hat{8}\hat{9}}\xi^{(0)}_{T4} =0
\label{mmcons}
\end{eqnarray}
The second and third constraints can be seen to be satisfied as in
the near horizon case by using an explicit representation of gamma
matrices. For now we concentrate on the $M6$ part. Using the first
constraint
 in (\ref{mmcons}), we see that the term containg
$(I+\tilde{\Gamma}_{\hat{0}\hat{1}})\xi$ in the kappa symmetry
matrix gives zero. Plugging in the values of vielbeins, we get, from
the remaining term, \be
\frac{1}{\sqrt{f}}(\sqrt{r^{2}+a^{2}}\cos\theta
\tilde{\Gamma}_{\hat{3}\hat{5}}+ r\sin\theta
\tilde{\Gamma}_{\hat{3}\hat{4}})(1-\tilde{\Gamma}
_{\hat{0}\hat{1}})\xi_{M6}^{(0)}= -\xi_{M6}^{(0)} \ee It is apparent
that the kappa symmetry condition cannot be satisfied for $r\not=0$.
For $r=0$ we get a projection condition on $\xi_{M6}^{(0)}$ that can
be easily seen to be inconsistent with the first of the constraints
in (\ref{mmcons}). We conclude that the D3 brane is not
supersymmetric in the full asymptotically flat geometry for any
value of $r$.

Let's ask why the supersymmetry of the D3 brane is broken in full
2-charge microstate geometry. We have seen that the Killing spinors
of the full six dimensional background geometry of $D3-D3$ system
satisfy the projection condition $(I +
\tilde{\Gamma}_{\hat{0}\hat{1}})\xi^{(0)}_{M6} =0$. In the near
horizon region, the geometry neatly separates into $AdS$ and sphere
parts, hence we can write gamma matrices
  for $AdS$ part and they act on the $AdS$ part of Killing spinor. So we have
\be \tilde{\Gamma}_{\hat{0}\hat{1}}\xi^{(0)}_{ads}= -\xi^{(0)}_{ads}
\label{adspr}\ee  In the near horizon region we have two types of
supersymmetries. In addition to ordinary supersymmetries, there are
also the  superconformal
  supersymmetries. Only ordinary supersymmetries continue to the
  far, i.e.,
asymptotically flat region.
   Now we want to see if the
projection condition (\ref{adspr}) is compatible with the kappa
symmetry condition for $D3$ brane wrapping the sphere in the near
horizon region. The condition one finds in the near-horizon region
is, with $\xi^{(0)}=\epsilon_{0}$, \be\label{eqn kappa condition 1}
\tilde{\Gamma}_{\hat{0}}\epsilon_{0}= - \epsilon_{0} \ee The
condition to be continuable to the far region is that it be in an
eigenvector of
$\tilde{\Gamma}_{\hat{0}\hat{1}}=\tilde{\Gamma}_{\hat{2}}$ i.e \be
\label{eqn kappa condition 2} \tilde{\Gamma}_{\hat{2}}\epsilon_{0} =
- \epsilon_{0} \ee

In three dimensions, $\tilde{\Gamma}_{\hat{0},\hat{1},\hat{2}}$ are
just Pauli matrices which don't commute. Hence it is not possible
for them
  to have simultaneous eigenvectors. As a result, the two conditions
(\ref{eqn kappa condition 1}) and (\ref{eqn kappa condition 2})
  are not compatible
and hence Killing spinors in the far region
 that are preserved by this $D3$ brane do not exist.
\newpage

\section{Discussion}
In \cite{Gaiotto:2004ij} the 4-charge black hole was considered. The charges
were D4-D4-D4-D0. It was argued that the D0 branes swell up into  D2
branes which wrap the horizon, and which occupies a Landau level on
the torus. The different ways to partition the D0 branes into such
groups gave the entropy of the hole.

We must however ask if the energy of the D0 branes remains the same
when we try to make them form a D2 brane; since we are looking at the
states of an extremal hole we do not have any `extra' energy to make
the D2 brane. It is not clear to us how this would work in general,
since in the limit where we have a very small D0 charge the mass of
the D2 brane would seem to be just the area of the horizon times the
tension, and this is much more than the mass of the D0 branes
attached to it. In fact in the work of \cite{Gaiotto:2004ij} the global
energy which follows from the supersymmetry algebra
turned out to be equal to the mass of the D2 brane {\em with
no contribution to the D0 charge}! This is what would follow
in our treatment if we chose a gauge for the 1-form potential of
the background to be $A^{(1)} = {R_{IIA}\over q_0}\sinh\chi~d\tau$
rather than (\ref{nesixteen}). As we have noted, there is always
an ambiguity in calculating energies from brane actions.

The situation would be clearer if we had an asymptotically flat
space-time.  With this in mind we have looked at 2-charge microstates
which have a similar structure to the system of \cite{Gaiotto:2004ij},
but where the AdS space inside goes over to asymptotically flat space
at large $r$. We find that the mass of the D3+P system (which is
analogous to the D2+D0 system) is given by the {\it sum} of two
contributions: the energy carried by P and an energy coming from the
tension of the D3.  It is interesting that the energy is given by such
a simple relation, because this configuration is not supersymmetric in
the full asymptotically flat geometry. This suggests that there is
some hidden symmetry in this 2-charge background, but we do not have
any clear understanding of this as yet.  But this also raises a puzzle
about the relation of this computation with that of
\cite{Gaiotto:2004ij}, since the mass of the D3+P system is more than
the mass of the P charge alone.

We also computed the gauge field produced by the D3 brane wrapped on
the $S^3$ in the full asymptotically flat geometry, and found that
the field strength went to a nonzero constant at infinity. This
suggests a divergent total energy for the field produced by the
brane, or alternatively, that the D3 branes and anti-branes wrapped
in this way are `confined' and cannot be separated to large distances
without generating a uniform energy density in the intervening
spacetime. Note that the energy $E=P+M$ computed using the DBI action
ignores this field energy.  The field energy is quadratic in the test
brane charge, and would be ignored in a linear analysis if it were
finite.

\section{Acknowledgements}

S.R.D. would like to thank the Tata Institute of Fundamental
Research for hospitality during the completion of this paper. The
work of S.R.D., X.W and C.Z was partially supported by the National
Science Foundation Grant No. PHY-0244811 and by the Department of
Energy Contract No. DE-FG-01-00ER45832. S.G. was supported by an
I.N.F.N. fellowship. We are grateful to B. de Wit, T. McLoughlin,
and J. Michelson for discussions.

\section{ Appendix :
Trajectories in the full Black Hole geometry}\label{fullbh}

Consider the motion of a brane in the full four dimensional black hole
geometry which has an energy (as measured in terms of the time in the
asymptotically flat region) which is given by $E=(M_2+M_0){R\over
q_0}$, i.e. the same energy which we found in the near-horizon
approximation. We will verify that this brane comes out of the horizon
and goes back and examine the parameter space for which the brane
remains in the near-hroizon region.  In this analysis we will set the
motion along the $T^6$ to zero from the beginning, so that we will
deal with the four dimensional part of the geometry.

The black hole solution is described in terms of harmonic functions
\ben
H_0(r) = 1 + {q_0 \over r}~~~~H_i(r) = 1 + {p_i \over r}~~(i=1,\cdots 3)
\label{jone}
\een
The (four dimensional part) string metric, dilaton and the 1-form
RR fields are given by
\bea
ds^2 & = & -{dt^2 \over [H(r)]^2} + [H(r)]^2~[dr^2 + r^2 d\Omega_2^2]
\nn \\
A_t & = & 1-{1\over H_0(r)} \nn \\
e^{\Phi} & = & {H_0(r) \over H(r)}
\label{jtwo}
\eea
where we have defined
\ben
H(r) = (H_0H_1H_2H_3)^{1\over 4}
\label{jthree}
\een
The lagrangian for a $D2$ brane which is wrapped on the $S^2$ at
some value of $r$ then becomes
\ben
S = -\mu (r) {\sqrt{[H(r)]^{-2}-[H(r)]^2 ({\dot{r}})^2}} + {M_0 \over
H_0(r)}
\label{jfour}
\een
where we have defined
\ben
\mu (r) = 4\pi \mu_2 {H(r)\over H_0(r)}{\sqrt{(H(r))^4r^4+f^2}}
\label{jfive}
\een
and the other quantities have been defined above.

The expression for the energy is
\ben
E = {\mu(r) [H(r)]^{-2} \over {\sqrt{[H(r)]^{-2}-[H(r)]^2
({\dot{r}})^2}}}
- {M_0 \over H_0(r)}
\label{jsix}
\een
Following the strategy of section (2.2) we will cast the problem
as that of a non-relativistic particle in some potential with the
non-relativistic energy equal to zero. The equation of motion
may be written using (\ref{jsix}) as
\ben
{1\over 2}({\dot{r}})^2 + W(r) = 0
\een
where
\ben
W(r) = - {1\over 2 H^2(r)}~ [ {1\over H^2(r)}- {\mu^2(r)
\over H^4(r) (E + {M_0 \over H_0 (r)})^2} ]
\label{jseven}
\een
The potential $W(r)$ behaves as $-r^4$ for small $r$ and $+r^4$
for large $r$ and has a single minimum. For any $E$ the brane
therefore starts from the horizon, goes upto a maximum distance
$r = r_0$ given by the point $W(r_0) = 0$ and turns back to the
horizon.

The near-horizon region has $r \ll q_0, p_i$ and we want to examine
whether $r_0$ lies in this region. The general problem is difficult
to analyze. However we get some indication by looking at the
simpler case where
\ben
q_0 = p_1 = p_2 = p_3 \equiv q
\een
so that $H_0(r)=H_1(r)=H_2(r)=H_3(r)=H(r)$. In this case
\ben
\mu^2 (r) = M_2^2 (1+{r \over q})^4 + M_0^2
\een
where $M_2$ is the $D2$ mass of the previous subsections.

\begin{figure}[ht]
\centerline{\epsfxsize=3.0in
\epsfysize=2.0in
    {\epsffile{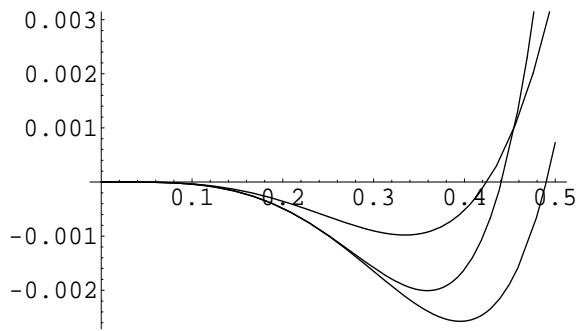}}
}
\caption{The potential $W(y)$ as a function of $y$ for $E=M_2$.
The curves have ${M_0\over M_2}=0,1,6$ starting from the top}
\label{full_geometry}
\end{figure}

In terms
of the dimensionless distance
\ben
y \equiv {r \over q}
\een
the potential $W(r)$ becomes
\ben
W(y)={y^4\over 2(1+y)^3}{y^2(1+y)^3-\epsilon^2 (1+y)-2\alpha\epsilon y
\over (\epsilon (1+y)+\alpha y)^2}
\label{jeighta}
\een
where we have defined
\ben
\epsilon \equiv {E\over  M_2}~~~~~~~~\alpha = {M_0 \over M_2}
\label{jnine}
\een
We want to examine only the special trajectory with $E=M_2$.
The function $W(y)$ for $E=M_2$ is shown in Figure
(\ref{full_geometry}) for various values of the ratio
$\alpha = M_0/M_2$

The trajectory will proceed to the zero of $W(y)$ at $y = y_0 (\alpha)
\neq 0$. The function $W(y)$ is plotted against $y$ for various values
of $\alpha$ in Figure (\ref{full_geometry}). It is clear that the
value of $y_0$ increases as $\alpha$ increases and becomes {\em greater
than unity} for sufficiently large $\alpha$. Thus the $D2$ brane
goes beyond the near-horizon region for large enough $\alpha$ and
strictly speaking the near-horizon approxiomation can be trusted
only when $M_0 \ll M_2$.

\end{document}